\documentclass[twocolumn]{aastex61}
\newcommand\pcc{\;{\rm cm}^{-3}}
\newcommand\Msun{\; M_{\odot}}
\newcommand\kms{\; {\rm km}\;{\rm s}^{-1}}

\newcommand\erg{\; {\rm erg}}

\newcommand\yr{\; {\rm yr}}

\newcommand\Myr{\;{\rm Myr}}

\newcommand\pc{\;{\rm pc}}
\newcommand\kpc{\;{\rm kpc}}

\newcommand\sfrunit{\Msun \kpc^{-2} \yr^{-1}}

\newcommand\Surf{\Msun\;{\rm pc^{-2}}}

\newcommand\Kel{\;{\rm K}}

\newcommand\simgt{\lower.5ex\hbox{$\; \buildrel > \over \sim \;$}}
\newcommand\simlt{\lower.5ex\hbox{$\; \buildrel < \over \sim \;$}}
\newcommand\pderiv[2]{\frac{\partial {#1}}{\partial {#2}}}

\newcommand\rbrackets[1]{\left({#1}\right)}
\newcommand\sbrackets[1]{\left[{#1}\right]}

\newcommand\abrackets[1]{\left\langle{#1}\right\rangle}
\newcommand\divergence[2][\rbrackets]{\nabla \cdot #1{#2}}

\newcommand\vel{\mathbf{v}}
\newcommand\Bvec{\mathbf{B}}

\newcommand\zhat{\hat{\mathbf{z}} }

\newcommand\Mflux[1]{\dot{\Sigma}_{{\rm wind}{#1}}}
\newcommand\Eflux[2][E]{\mathcal{F}_{{#1}{#2}}}
\newcommand\vout{v_{\rm out}}
\shorttitle{Galactic Winds and Fountains}
\shortauthors{Kim \& Ostriker}

\begin{document}

\title{Numerical Simulations of Multiphase 
  Winds and Fountains from Star-Forming Galactic Disks: I.
  Solar Neighborhood {\it TIGRESS} Model}

\author[0000-0003-2896-3725]{Chang-Goo Kim}
\affiliation{Department of Astrophysical Sciences, Princeton University, Princeton, NJ 08544, USA}
\affiliation{Center for Computational Astrophysics, Flatiron Institute, New York, NY 10010, USA}
\author[0000-0002-0509-9113]{Eve C. Ostriker}
\affiliation{Department of Astrophysical Sciences, Princeton University, Princeton, NJ 08544, USA}
\email{cgkim@astro.princeton.edu, eco@astro.princeton.edu}

\begin{abstract}
  Gas blown away from galactic disks by supernova (SN) feedback plays a
  key role in galaxy evolution.
  We investigate outflows utilizing the solar
  neighborhood model of our high-resolution, local galactic disk
  simulation suite, TIGRESS.  In our numerical implementation, star
  formation and SN feedback are self-consistently treated and well
  resolved in the multiphase, turbulent, magnetized interstellar
  medium.  Bursts of star formation produce spatially and temporally
  correlated SNe that drive strong outflows, consisting of hot
  ($T>5\times10^5\Kel$) winds and warm ($5050\Kel < T <
  2\times10^4\Kel$) fountains.  The hot gas at distance $d>1\kpc$ from
  the midplane has mass and energy fluxes nearly constant with $d$.
  The hot flow escapes our local Cartesian box barely affected by
  gravity, and is expected to accelerate up to terminal velocity of
  $v_{\rm wind}\sim350-500\kms$.  The {mean}
  mass and energy loading factors
  of the hot wind are 0.1 and 0.02,
  respectively.  For warm gas, the {mean} outward mass
  flux through $d=1\kpc$ is comparable to the mean star formation
  rate, but only a small fraction of this gas is at velocity  $>50\kms$.
  Thus, the warm outflows eventually fall back as inflows.
  The warm fountain flows are created by expanding hot superbubbles
  at $d\simlt 1\kpc$; at larger $d$ neither ram pressure acceleration
  nor cooling transfers significant momentum {or energy} flux from the hot wind to the
  warm outflow.  The velocity distribution at launching {near $d\sim1\kpc$} better represents
  warm outflows than a single mass loading
  factor, potentially enabling development of subgrid models for warm
  galactic winds in arbitrary large-scale galactic potentials.
  \end{abstract}

\section{INTRODUCTION}\label{sec:intro}

Galactic scale gas outflows (or winds) are ubiquitous in star forming
galaxies \citep[see][for
  reviews]{2005ARA&A..43..769V,2017arXiv170109062H} and believed to be
essential to distribution of the gas and metals in
galaxies and the circumgalactic/intergalactic medium (CGM/IGM) and hence
to regulating cosmic star formation history 
\citep[see][for reviews]{2015ARA&A..53...51S,2017ARA&A..55...59N}.
Theoretical models of the stellar mass-halo mass relation 
constructed by abundance matching of
observational stellar mass functions to
simulated halo mass functions
\citep[e.g.,][]{2013MNRAS.428.3121M,2013ApJ...770...57B,2017MNRAS.470..651R}
indicate 
that galaxies (or dark matter halos) are very inefficient
at converting gas into stars.  At low redshift, at most
$\sim10-20\%$ of the available
baryonic mass has been converted into stars at halo mass of
$\sim10^{12}\Msun$, while the ratio of stellar mass to halo mass
declines steeply toward both lower and higher masses.  Recent
cosmological hydrodynamic simulations of large volumes of the Universe
require strong outflows driven by both star formation
and active galactic nuclei feedback to explain low baryonic abundance
in galaxies compared to the cosmic fraction
\citep[e.g.,][]{2003MNRAS.339..289S,2014Natur.509..177V,2015MNRAS.446..521S}.

Direct inclusion of feedback processes in large-volume cosmological
galaxy formation simulations is still not feasible in practice.
For star formation feedback by supernovae (SNe), 
implementation via simple thermal energy dumps suffers 
``overcooling,'' with energy radiated away without 
preventing in situ star formation or driving winds 
\citep[e.g.,][]{1992ApJ...391..502K}; this is because resolving
the cooling radii of SN remnants requires much higher resolution 
\citep{2015ApJ...802...99K} than is practicable in large-volume
simulations.  In cosmological zoom-in simulations, 
more careful treatments of SNe allowing for a
``momentum prescription'' at low resolution 
can solve at least some aspects of the overcooling problem 
\citep[e.g.,][]{2014ApJ...788..121K,2015MNRAS.451.2900K,
  2014MNRAS.445..581H,2017arXiv170206148H}, especially for dwarfs.
However, given the constraints of computational expense,
treating unresolved physics with parameterized models
is unavoidable in many situations, 
including in simulations of galaxy groups/clusters, and
in the large boxes needed for
fully-sampled statistics of cosmic galaxy populations
\citep[e.g.,][]{2015MNRAS.446..521S,2017arXiv170703406P}. 

When star formation feedback physics is not directly simulated,
galactic winds are not an outcome but an input that is
part of the  ``subgrid'' treatment  \citep{2015ARA&A..53...51S}. 
Currently, however,
subgrid models of wind driving by stellar feedback
often either adopt highly simplified scaling
prescriptions for wind mass loss rates (relative to the star formation rate)
and velocities (relative to the halo potential depth), or else are 
calibrated using empirical results from a limited set of galaxies
(and hence are not fully predictive). 
Better theoretical models are clearly needed.  
Towards this end, the first step is to provide a physical
understanding and detailed characterization of outflowing gas (including
both winds and fountains) in
galaxies, informed and calibrated based on high-resolution
three-dimensional numerical magnetohydrodynamic (MHD) simulations.
To fully capture the
interaction between stellar feedback and a realistic multiphase interstellar
medium (ISM), it is crucial to self-consistently include the gravitational
collapse that produces star clusters and to resolve the local injection of
energy from individual massive stars.  

In classical theoretical models of galactic winds motivated by observed
starburst galaxies \citep[e.g.][]{1985Natur.317...44C}, a
steady, adiabatic flow is produced
by a central energy source. In this approach,
hot, overpressured gas flows are characterized by ``mass loading''
and ``energy loading''
factors, respectively defined by the ratios of mass and energy outflow rates
to star formation rates and energy injection rates
at the wind base.  Simple spherical wind models can also be constructed
that allow for radiative cooling, such that the temperature precipitously
drops at some radius in certain parameter regimes 
\citep[e.g.][]{1995ApJ...444..590W,2016ApJ...819...29B,2016MNRAS.455.1830T}.

Observed galactic outflows are multiphase in nature. 
Systematic observations reveal 
prevalent multiphase structure of galactic winds with cold molecular
\citep[e.g.,][]{1999A&A...345L..23W,2015ApJ...814...83L}, neutral \citep[e.g.,][]{2000ApJS..129..493H,2005ApJ...621..227M,2005ApJS..160..115R,2010AJ....140..445C,2013A&A...549A.118C},
ionized \citep[e.g.,][]{2001ApJ...554..981P,2003ApJ...588...65S,2010ApJ...717..289S,2012ApJ...759...26E,2015ApJ...809..147H,2017MNRAS.469.4831C}, and hot gas phases \citep[e.g.,][]{2000MNRAS.314..511S,2007ApJ...658..258S,2017MNRAS.467.4951L}.
For the best studied example, local starburst M82, \citet[][see also \citealt{2016ApJ...830...72C}]{2015ApJ...814...83L}
have shown a clear signature of decreasing {outward}
 mass fluxes in molecular and neutral gas
as a function of the distance from the disk midplane, implying a fountain
\citep{1976ApJ...205..762S,1980ApJ...236..577B} rather
than a wind
for the cooler gas. 
In one conceptual framework, warm and cold gas in outflows 
results when a hot medium accelerated by its
own pressure gradient cools radiatively; an alternative concept is
that overdense warm and cold
ISM clouds are ``entrained'' by a high-velocity, low-density
hot wind.  More realistically, both effects can in principle
occur, and in general there is
a complex interaction between the multiple phases that are present.

The mass and energy loading factors are key quantities that
characterize winds and quantify their significance in controlling
baryonic mass cycles of galaxies. Measuring these loading factors has
been of intense observational interest, but uncertainties are still
large. In particular, the reported mass loading factor ranges widely
from 0.01 to 10 \citep{2005ARA&A..43..769V}.  Depending on the
assumed geometry, metallicity, and ionization state, the mass outflow rate
can easily be reduced by a factor of 10
\citep[e.g.,][]{2016MNRAS.463..541C,2017MNRAS.469.4831C}.  In
addition, uncertain deprojection may result in an
overestimate the velocity, incorrectly leading to
interpretation of a low-temperature outflow as a wind rather than a fountain.
If gas is not really escaping, the {outward} mass flux will be a decreasing
function of distance from the wind launching region, and mass fluxes
estimated at small radii would exceed the true losses from a galaxy.

Predicting wind loading factors theoretically requires
modeling the interaction between SN remnants (including
from clustered SNe) and the ISM.
Expansion of superbubbles driven by multiple SNe
has been studied by idealized analytic models and simulations 
\citep[e.g.,][]{
  1987ApJ...317..190M,1988ApJ...324..776M,
  1989ApJ...337..141M,1999ApJ...513..142M,2017ApJ...834...25K}.
While these idealized models provide essential physical insight
and quantitative estimates,
firm theoretical measurements of mass loading in multiphase winds
from galactic disks require ISM models with realistic spatio-temporal
distribution of SNe and vertical stratification.
A number of local stratified-disk
simulations, with increasingly high resolution, have modeled
the multiphase ISM with SN feedback  
\citep[e.g.,][]{1999ApJ...514L..99K,2000MNRAS.315..479D,
  2004A&A...425..899D,2006ApJ...653.1266J,2012ApJ...750..104H,
  2013MNRAS.432.1396G,2015MNRAS.454..238W,2016MNRAS.456.3432G,
  2017ApJ...841..101L}, albeit with SN rates and locations imposed 
rather responding self-consistently to star formation.
Some recent numerical work has focused specifically on the outflow properties
driven by SN feedback 
\citep[e.g.,][]{2013MNRAS.429.1922C,2016MNRAS.459.2311M,2017MNRAS.470L..39F}, 
although with a cooling cutoff at $10^4\Kel$ that does not allow for
the full range of ISM phases.

Very recently, it has become possible to evolve the turbulent, magnetized,
multiphase ISM in local galactic disks with cooling and heating, 
galactic differential rotation, and self-gravity, including
fully self-consistent resolved star formation 
and SN feedback over durations of several 100 Myr
\citep[][Paper~I hereafter]{2017ApJ...846..133K}. 
A few other recent simulations have also included self-gravity to model
SN rates and positions self-consistently with star formation
\citep[][]{2014A&A...570A..81H,2017MNRAS.466.1903G,2017MNRAS.466.3293P,2017A&A...604A..70I},
but given their relatively short simulation duration ($\le 100\Myr$),
they have not reached a statistically quasi-steady state and wind properties
may be subject to transient effects from the simulation start-up.

In this paper, we analyze our fiducial model from 
the TIGRESS (Three-phase ISM in Galaxies Resolving Evolution with Star formation
and SN feedback) simulation suite introduced in Paper~I, in order to provide
more comprehensive understanding of multiphase gas outflows
in the realistic ISM. 
Our analysis here mainly
focuses on characterizing differences among outflows of
different thermal phases.
In a subsequent paper, we will analyze
models with different galactic conditions 
to investigate systematic scaling relations of wind properties 
\citep[e.g.,][]{2015MNRAS.454.2691M,2015ApJ...809..147H,
2016ApJ...822....9H,2017MNRAS.469.4831C}.

In Section~\ref{sec:overview},
we review equations for steady, adiabatic flows 
and summarize key physical quantities to be measured from the simulation.
We then present an overview of gas flows in the simulation, including overall
mass fluxes and vertical profiles of each gas phase; this demonstrates
the necessity of a phase-by-phase analysis.
In Section~\ref{sec:wind}, we analyze the hot gas component, showing
that it is consistent with a wind having
well-defined  mass flux and specific energy 
(or Bernoulli parameter) that are {approximately} constant
as a function of distance from the midplane.
Section~\ref{sec:fountain} presents an analysis of the warm gas component,
showing characteristics of a fountain flow that has 
decreasing mass and energy fluxes
as a function of distance from the midplane.
Section~\ref{sec:loading} provides
mass and energy loading factors of each phase, comparing these to previous work 
and to observations.
Section~\ref{sec:summary} summarizes our main conclusions.

\section{Overview of Gas Flows in TIGRESS}\label{sec:overview}

{\subsection{Outflow terminology, vertical profiles, and classical adiabatic winds}}

In galactic disks, star formation takes place in dense gas
near the midplane, within the scale height of the ISM.
Prodigious energy is injected by SNe within this thin layer,
and the high-entropy, overpressured gas expands outward.
Strong shocks heat and accelerate both dense, cold cloudlets and the warm,
diffuse intercloud medium surrounding individual SNRs and superbubbles,
with most of this gas cooling back to its original temperature relatively
quickly \citep[e.g.][]{2015ApJ...802...99K,2017ApJ...834...25K}.
Depending on the level of remaining specific energy with respect to the 
gravitational potential, outflows {of cooled, SN-accelerated warm
  (or cold) gas may either keep moving out of the disk,
  or may turn around at some height.
  While most of the energy deposited by SNe is radiated away,
  some of the hot gas created in strong SN shocks is at low enough
  density that it has very limited cooling.  This accelerates away from
  the midplane towards higher-latitude, lower-pressure regions, achieving
  high enough velocity that it can escape from the galactic potential well.}
In this paper, the term  ``galactic winds'' refers to outflows
launched with high enough energy to escape the
galactic gravitational potential, while the term ``galactic fountains''
refers to outflows launched with insufficient energy that eventually fall back.

{ In our simulations (and in real galaxies), the motions of gas
  are three-dimensional.  Within any given patch of the ISM in the disk, followed
  on its orbit about the galactic center, horizontal averaging can be
  used to define a mean density, velocity,  and other flow properties
  as a function of height $z$.  In general, the instantaneous mean velocity 
  will have both horizontal (radial-azimuthal) and vertical
  components. The horizontally-averaged flow quantities may be further
  averaged over time (with window comparable to an orbit time, so that
  epicyclic motions are averaged away).  If the accretion rate is low,
  the resulting temporally-averaged velocity at any height
  will be dominated by vertical motion.
  Thus, horizontal- and temporal-averaging 
  of the gas yields an effectively one-dimensional profile
  as a function of height, consisting of average values
  $\langle \rho(z)\rangle$, $\langle v_z(z)\rangle$, $\langle P(z) \rangle$,
  etc.    If there is a net outflow $sign(v_z)=sign(z)$, and if
  there is a net inflow the sign is reversed.}  

{Similar to horizontal- and time-averaged flows, classical SN-driven
  wind solutions are one-dimensional, and the simplest solutions are also adiabatic.} 
For steady-state one-dimensional adiabatic gas flows,
the equations of mass and total energy conservation including source terms
can be written as
\begin{equation}\label{eq:mass}
\divergence{\rho \vel}=\dot{\rho}_{\rm inj}(z),
\end{equation}
\begin{equation}\label{eq:energy}
\divergence{\rho \vel\mathcal{B}}=\dot{e}_{\rm inj}(z),
\end{equation}
where $\dot{\rho}_{\rm inj}$ and $\dot{e}_{\rm inj}$
are the volumetric mass and energy injection rates, respectively,
arising from SN feedback.
{In Equations~(\ref{eq:mass}) and (\ref{eq:energy}), $z$ represents the
  vertical direction for a flow perpendicular to the plane of a galactic disk;
  an approximately spherical galactic center 
  flow \citep[e.g.][]{1985Natur.317...44C} would
  instead have $\dot{\rho}_{\rm inj}(r)$ and $\dot{e}_{\rm inj}(r)$.
  Note that $\dot{\rho}_{\rm inj}$ does {\it not} represent SN ejecta itself. 
  Rather, shock heating of ambient ISM gas near SNe increases the
  mass injection rate above that of 
  the initial SN ejecta, while cooling tends to reduce the rate.
  Allowing for this shock heating and cooling,
  $\dot{\rho}_{\rm inj}$ is simply the mean local rate
  at which hot material is added the steady flow.
  Similarly, $\dot{e}_{\rm inj}$ represents the 
  mean local energy input rate to the flow,
  which is bounded above by the initial
  energy carried by SN ejecta.} 

In Equation (\ref{eq:energy}), the
total specific energy (the Bernoulli parameter)
is defined by 
\begin{equation}\label{eq:ber}
\mathcal{B}\equiv\frac{v^2}{2}+\frac{\gamma}{\gamma-1}\frac{P}{\rho}+\Phi,
\end{equation}
consisting of specific kinetic energy, specific enthalpy
$h=[\gamma/(\gamma -1)]P/\rho$, 
and gravitational potential terms.  
Note that here, for simplicity, we consider adiabatic, unmagnetized gas, but
Appendix~\ref{sec:appendix} presents {
the full equations for general 
MHD flows in a local shearing box, and shows that
horizontal- and time-averaging yields a set of 
simple steady-state 1D flow equations, which can be
applied} to our simulations.  
In this paper, the $\Phi=0$ reference point is at the midplane.

For flows in a local Cartesian box, like ours, 
with energy sources near the midplane,
{any time-averaged steady winds that may exist }
are launched vertically along the $\hat z$ axis. 
Through Gauss's Law, volume integration of Equations~(\ref{eq:mass})
and (\ref{eq:energy}) gives 
mass and total energy fluxes 
through surface area $A=L_xL_y$ at
$z=\pm d$ 
as
\begin{equation}\label{eq:Mflux}
\Mflux{}(d)\equiv F_M(z=d)-F_M(z=-d)=\dot{M}_{\rm inj}(d)/A
\end{equation}
and
\begin{equation}\label{eq:Eflux}
\Eflux{}(d)\equiv F_E(z=d)-F_E(z=-d)=\dot{E}_{\rm inj}(d)/A.
\end{equation}
Here, $d$ is the distance from the disk midplane,
and total mass and energy injection rates within $|z|<d$ are
$\dot{M}_{\rm inj}(d)=\int_{-d}^{d} \dot{\rho}_{\rm inj} dV$ 
and $\dot{E}_{\rm inj}(d)=\int_{-d}^{d} \dot{e}_{\rm inj} dV$.
The quantities $F_M(z)=\langle \rho v_z \rangle$
and $F_E(z)=\langle \rho v_z\mathcal{B}\rangle$
stand for mass and total energy fluxes 
averaged over horizontal area at height $z$.\footnote{
  For the energy flux, we hereafter use subscripts KE, TE, GE, and ME
to denote different energy components.  These are respectively  
$F_{\rm KE}=\rho v_z v^2/2$ (kinetic), $F_{\rm TE}=\rho v_z h$ (thermal,
with $h$ enthalpy), 
$F_{\rm GE}=\rho v_z \Phi $ (gravitational),
and $F_{\rm ME}=S_z$ (magnetic;
see Section~\ref{sec:appendix} and 
Equation (\ref{eq:a_Poynting_z})).   
The subscript E will denote the total energy term, the sum of all four
components.}
The above relations assume periodic boundary conditions in both of the
horizontal directions; energy terms associated with the background shear from
integrals over faces perpendicular to $\hat x$ are discussed in
Appendix~\ref{sec:appendix}.

Since SN explosions are usually concentrated within a thin layer near the midplane, 
$\dot{M}_{\rm inj}(d)$ and $\dot{E}_{\rm inj}(d)$ are expected to approach
constant values for $d\gg H$, where H is the disk scale height. 
Hence,
if $A$ is constant as in the local Cartesian coordinates, 
for a steady wind the (areal) mass and energy fluxes $\Mflux{}$ and $\Eflux{}$ 
would also approach constants at $d\gg H$ above the source region.
For general geometries (e.g., \citealt{1985Natur.317...44C} 
for spherical coordinates), however,
where the area that encloses a given set of
streamlines varies as a function of distance (e.g., $A\propto r^2$ for the
spherical case), 
the fluxes would also vary (e.g., $\propto r^{-2}$).

To the extent that horizontal correlations (or anticorrelations)
among variables at a given $z$ may be neglected,
$F_E=F_M \mathcal{B}$, and if we may assume symmetry across the midplane
this yields $\mathcal{B} = \mathcal{F}_E/\dot{\Sigma}_{\rm wind}=\dot{E}_{\rm inj}/\dot{M}_{\rm inj}$.   
This implies that beyond the
source region where $\dot{E}_{\rm inj}$ and $\dot{M}_{\rm inj}$ have reached
their final values, the Bernoulli parameter becomes constant
{for a steady (or time-averaged) flow.}  
In the more general case with $A$ an increasing function of distance,
because the area perpendicular to streamlines is the same for both mass and
energy flux, $F_E$ and $F_M$ would vary {$\propto A^{-1}$} but 
$\mathcal{B}$ would still be conserved along
streamlines (beyond the source region) irrespective of geometry.  Therefore,
for steady {adiabatic  winds (or equivalently
for any time-averaged, adiabatic  portion in 
a more general outflow),}
the Bernoulli parameter is a key quantity
that enables robust extrapolation 
of flow evolution to large distances.\footnote{This holds true for
  MHD flows if the Poynting flux is negligible, 
 as in our simulations; see Figure~\ref{fig:hot}.}
As applied to the present problem,
this suggests that evaluation of $\mathcal{B}$ with our local disk simulations
should provide predictions for properties {of the hot portion of the wind} 
at large distance
(outside our simulation domain) where
wind streamlines open up (becoming more radial than vertical).
This motivates the need to quantify mass and energy fluxes
in the launching region -- just above the source region -- in our simulations.

The simple analysis above provides helpful intuition for gas flows driven
by localized energy sources, but the real ISM -- and our simulations -- is not 
a single phase, adiabatic gas.  In fact, material in the ISM spans a  
wide range of density and temperature. 
SN shocks are responsible for
generating the hot gas phase ($T\sim 10^6-10^7\Kel$), 
which interacts with surrounding warm ($T\sim 10^4\Kel$) 
and cold ($T\sim 10^2\Kel$) phases.
Considering each thermal component
\textit{individually}, radiative heating and cooling as
well as mass and energy transfers between components  
would act as source and sink terms in the conservation equations of each phase. 
Because SN ejecta strongly interact with the surrounding
ISM in extremely
complex ways to heat and accelerate gas, some of which may be able to
escape from a galaxy, it is not at all obvious how one would estimate
$\dot{M}_{\rm inj}$ and $\dot{E}_{\rm inj}$
{for individual thermal components of a multiphase outflow.}
Moreover, star formation and hence SN events 
are very bursty, and this burstiness may affect yields.
Clearly, {the total and individual-phase}
$\dot{M}_{\rm inj}$ and $\dot{E}_{\rm inj}$ are 
only quantifiable with self-consistent numerical simulations that capture
the full physics of the ISM.
{ Nevertheless, while simulations are essential for obtaining the detailed properties of
  realistic multiphase outflows, we can
  still expect certain aspects of classical wind solutions to hold when
  suitably applied.  In particular, as we shall
  show, the space-time-averaged hot portion of the wind, when considered
  separately from other phases, shares many
  similarities with adiabatic one-dimensional winds. }

{\subsection{TIGRESS simulation model and analysis}}

In Paper~I, we presented a novel framework for
multi-physics numerical simulations of the star-forming 
ISM implemented in the {\it Athena} MHD code
\citep{2008ApJS..178..137S,2009NewA...14..139S}. 
We solve the ideal  MHD equations in a local, shearing
box, representing a small patch of a differentially rotating
galactic disk. This treatment allows us to
achieve uniformly
high spatial resolution compared to what is possible in
a global simulation of an entire
galaxy (or galaxies) \citep[e.g.,][]{2012MNRAS.421.3522H,2015MNRAS.454.2691M}, 
which is crucial to resolve both star formation and SN feedback
as well as all thermal phases of the ISM both near and far from the midplane. 
We include gaseous and (young) stellar self-gravity, a fixed external
gravitational potential to represent the old stellar disk and dark matter halo,
galactic differential rotation, and optically thin cooling and grain
photoelectric heating. We utilize sink particles to follow formation of and
accretion onto star clusters in dense, cold gas. Massive young stars in these 
star clusters feed energy back to the ISM, by emitting far-ultraviolet
radiation (FUV) and exploding as SNe. The former heats the diffuse
warm and cold ISM, while the latter creates hot ISM gas, drives turbulence, and
induces outflows.

Our simulations yield
realistic, fully self-consistent three-phase ISM models with self-regulated
star formation.\footnote{{Of course, the absence of global geometry
  means that we are unable to follow effects of strong noncircular flows in
  the disk, transport of gas from one radius to another in a fountain
  (which would also require significant angular momentum exchange), or
  the transition of hot winds through a sonic point.  Nevertheless, the
  high resolution afforded by our local scope is extremely valuable for
  limiting artificial mixing, which is essential for understanding
  key characteristics of multiphase flows}.}
  Paper~I presented results from a fiducial model with
parameters similar to those of the Solar neighborhood. There, we showed that
after $t\sim100\Myr$ a quasi-steady state is reached.
When stars form, massive stars
enhance heating in the warm and cold ISM,
and the SN rate increases, driving turbulence throughout the ISM. 
Both feedback processes disperse dense clouds and puff the gas disk up,
{temporarily}
shutting star formation off. With the corresponding reduction in
star formation feedback, the gas can settle back to the midplane and
collect into dense clouds which then form a new generation of stars.
In Paper~I, we evaluated several basic ISM and wind properties, and 
demonstrated their convergence as a function of the numerical spatial resolution.

\begin{figure*}
\plottwo{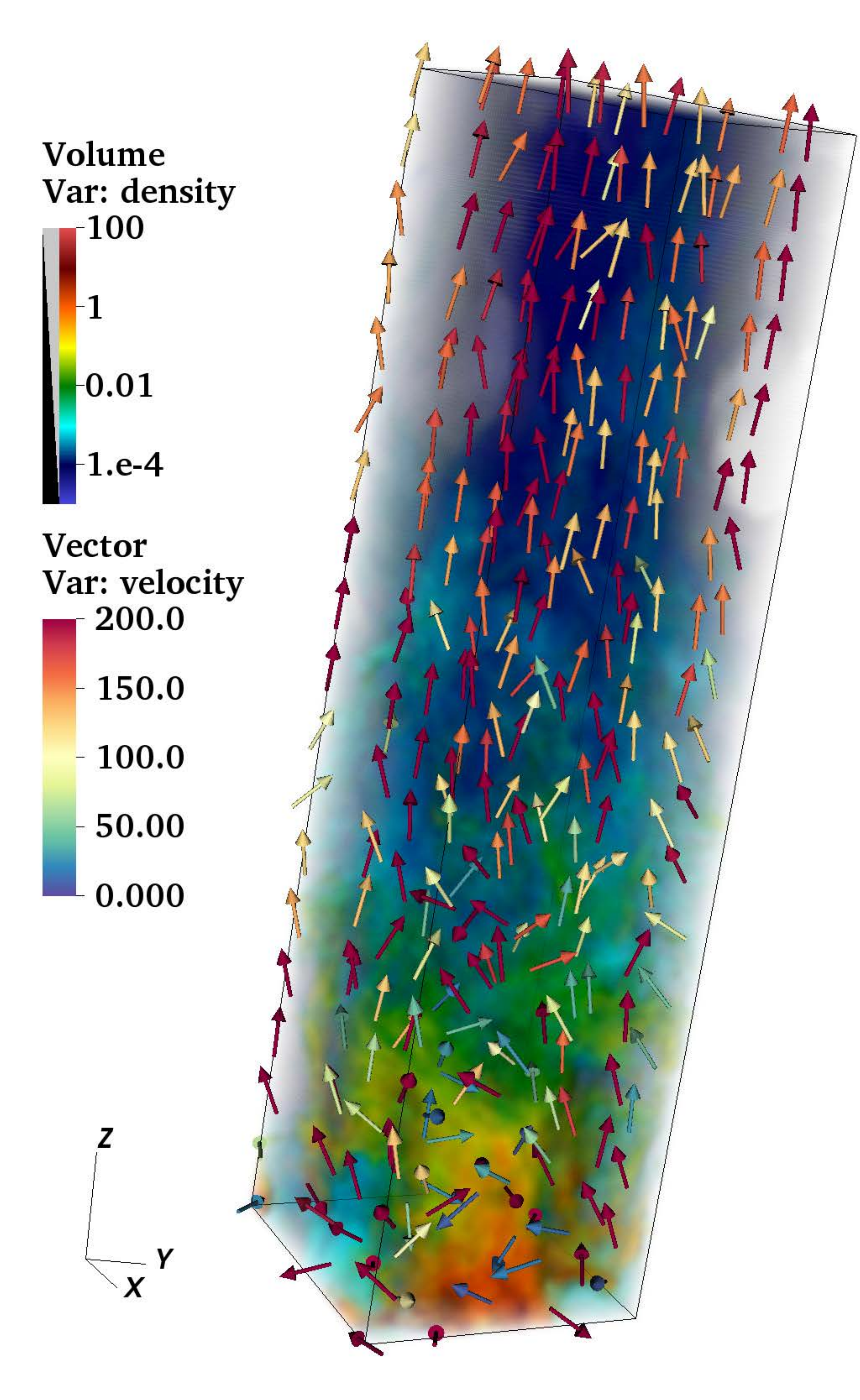}{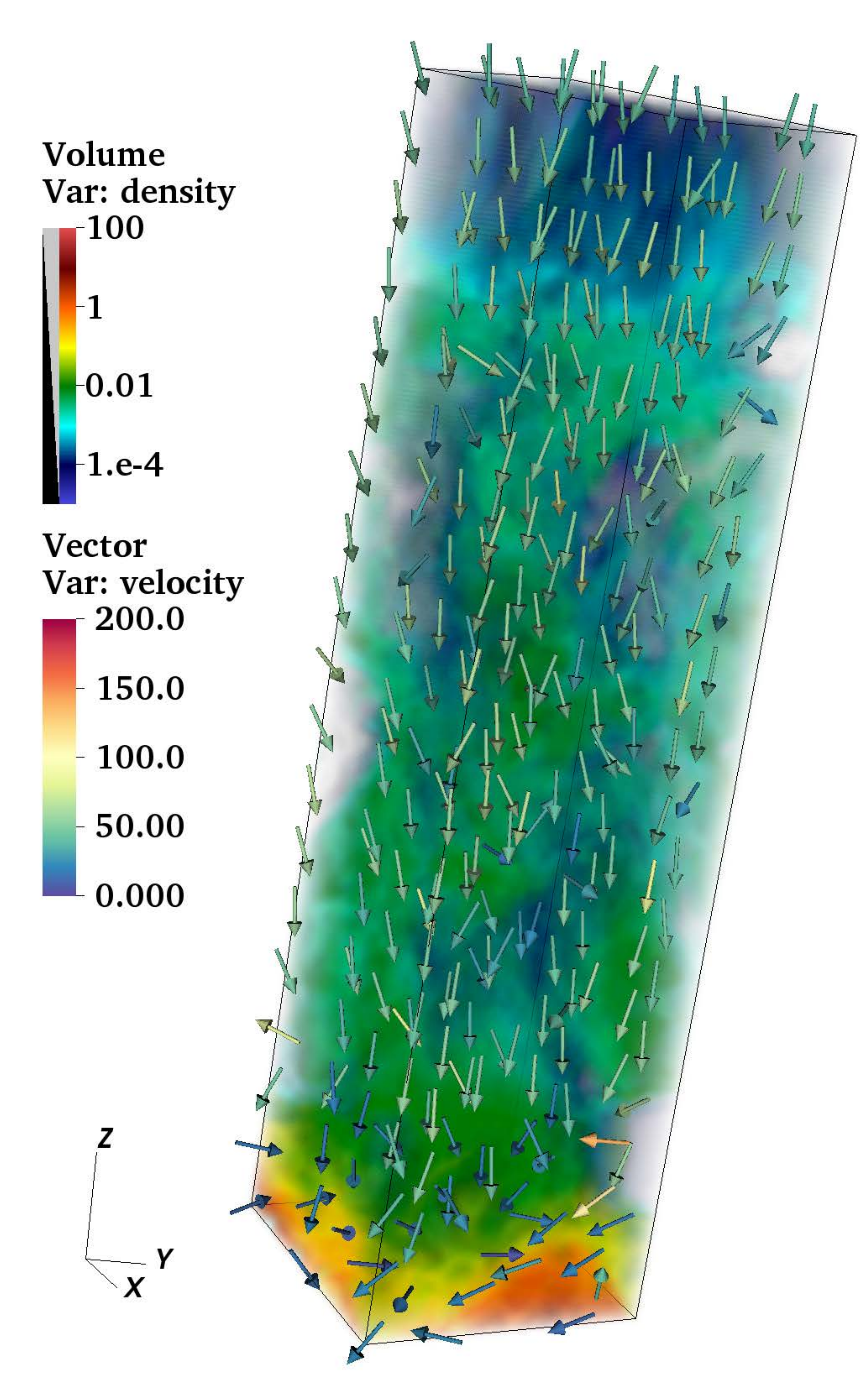}
\caption{Sample snapshots illustrating 
  (left) outflow-dominated,  and (right) inflow-dominated periods, 
  at $t=300$ and $360\Myr$, respectively.  Gas density is shown in  
  color scale volume rendering and the velocity field is shown with
vectors; vector colors (rather than length)
indicate velocity magnitudes. 
A fast-moving, low-density (dark blue) outflow is evident in the left snapshot, 
while moderate-density (green) gas that was previously
blown out to large distances is falling back toward the midplane in the right.
Velocity fields are turbulent near the midplane ($d<H$), but ordered
in either outflowing or inflowing directions at large distances.
For visual clarity, only the upper half of the simulation box
($z=0$ to $4\kpc$) is shown, with a full horizontal crossection
$-512\pc \le x,y \le 512\pc$.  
\label{fig:snap}}
\end{figure*}

\begin{figure}
\plottwo{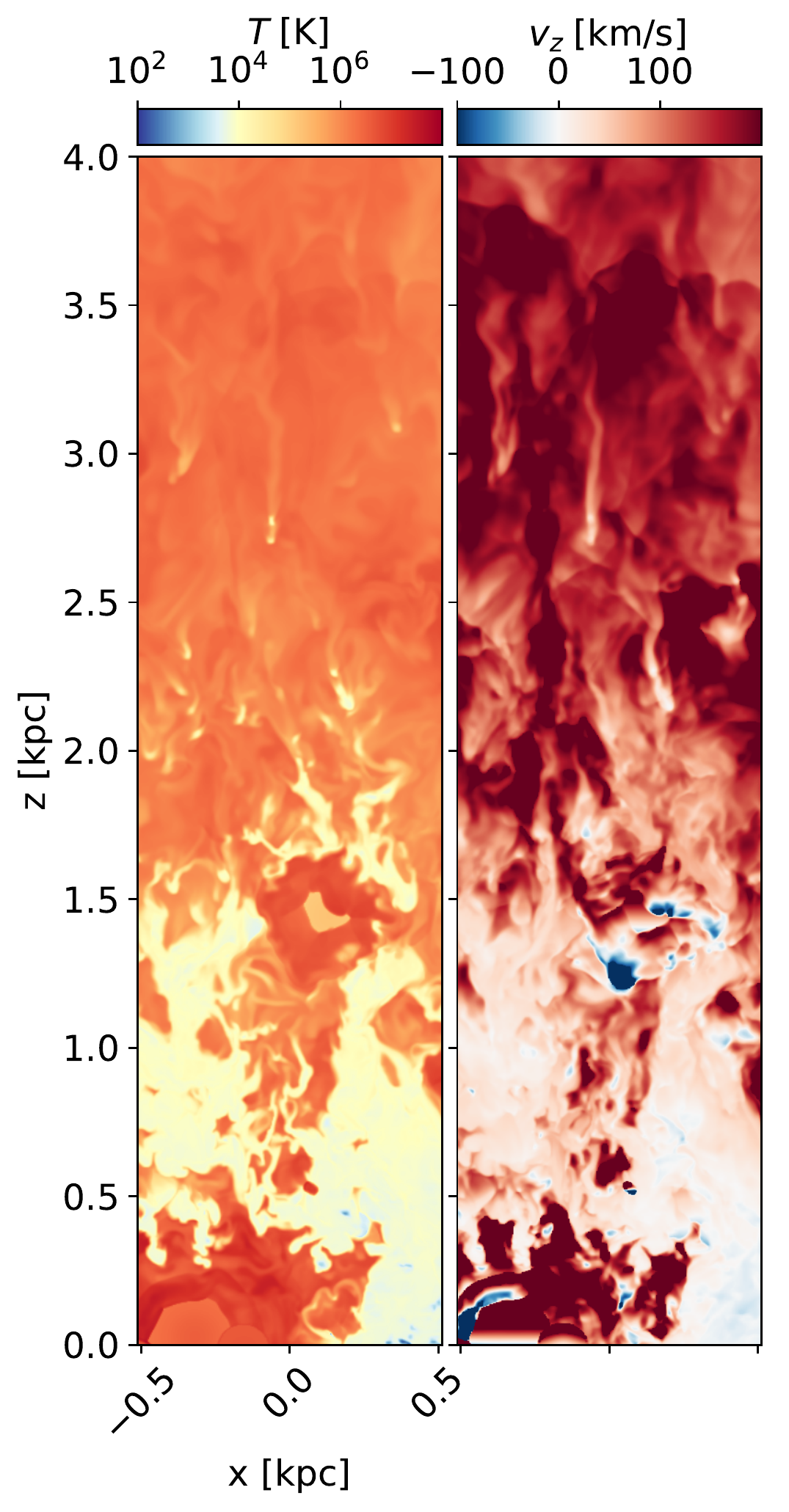}{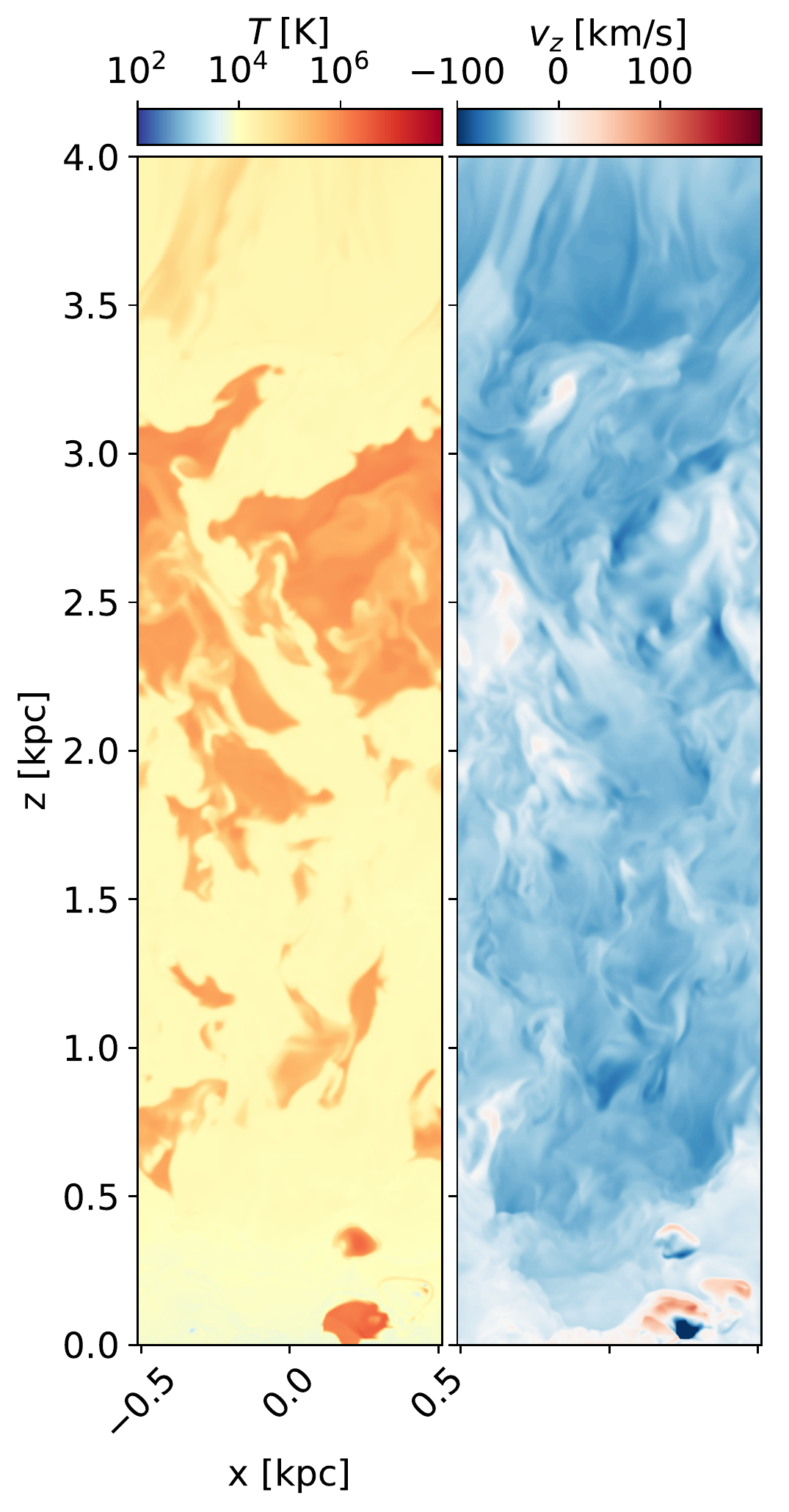}
\caption{Sample slices through $y=0$ showing temperature and vertical velocity
at $t=300$ (left) and $360\Myr$ (right), 
representing outflow- and inflow-dominated periods.
During outflow-dominated periods (left), the hot component
fills most of the volume at high $|z|$ and flows outward at high velocity,
while the warm component is confined in small cloudlets.
During inflow-dominated periods (right),
the warm component occupies most of the volume and falls toward the midplane.
As in the volume renderings of Fig. \ref{fig:snap},
only the upper half of the simulation box ($z=0$ to $4\kpc$) is shown.  
\label{fig:slice}}
\end{figure}

\begin{figure}
\plotone{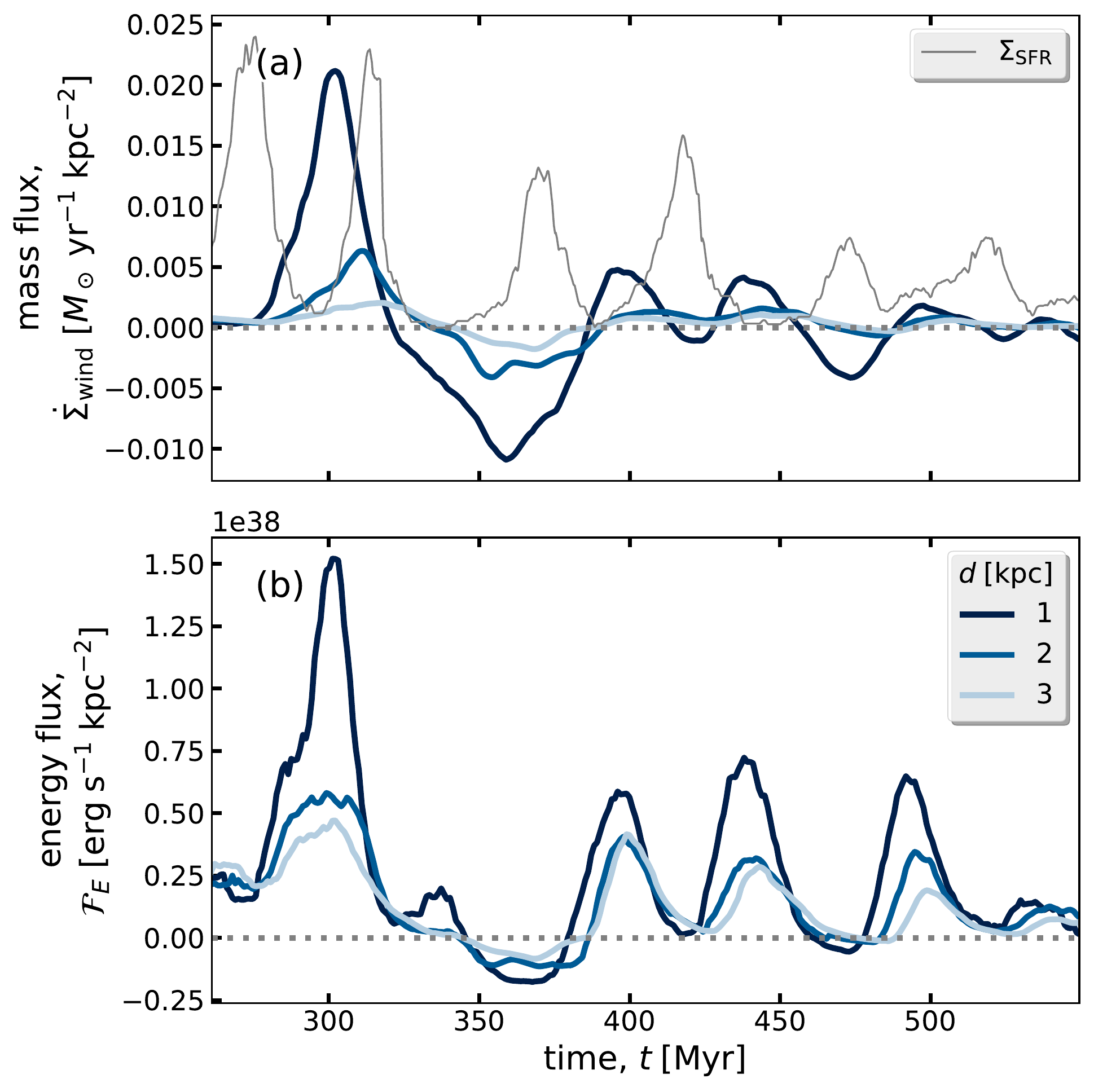}
\caption{
Time evolution of (a) mass and (b) energy fluxes driven by 
bursts in the star formation rate.
Mass and energy fluxes per unit area at distances 
$d=$1, 2, and 3 kpc from the midplane are measured 
(Equations~\ref{eq:Mflux} and \ref{eq:Eflux}).
Each burst in $\Sigma_{\rm SFR}$ (shown in grey in upper panel)
leads to outflows in both mass and energy near the midplane.
The net mass flux decreases at larger $d$ because most of the outflow near
the midplane is warm ``fountain'' gas with low velocity that turns around
{(resulting in periods of negative mass flux at $d=1\kpc$).}  
Energy fluxes are substantial at all heights, due to the dominance of
hot gas that escapes as a wind. 
\label{fig:flux_tevol}}
\end{figure}

In this paper, we analyze detailed properties of galactic winds and fountains
for a vertically extended version of the fiducial model of Paper~I.  
The simulation domain covers 
$1\kpc\times1\kpc$ horizontally and $-4.5\kpc<z<4.5\kpc$ vertically, at a 
uniform resolution $\Delta x=4\pc$.
Representative snapshots displaying a
volume rendering of density and velocity vectors
during outflow- and inflow-dominated periods are shown in
Figure~\ref{fig:snap}.
Figure~\ref{fig:slice} displays slices of temperature and vertical velocity
through the $y=0$ plane for the same snapshots shown in Figure~\ref{fig:snap}.
The outflows and inflows seen in Figures~\ref{fig:snap} and \ref{fig:slice}
are part of an overall
cycle that repeats, representing the response to large amplitude
temporal fluctuations in star formation rates 
(see grey line in Figure~\ref{fig:flux_tevol}(a)).

To characterize vertical gas flows, we first construct 
horizontally-averaged quantities. We then calculate the {net}
mass and energy fluxes
that pass through horizontal planes (both upper and lower sides) 
at $d=1$, $2$ and $3\kpc$ 
(cf. Equations~\ref{eq:Mflux} and \ref{eq:Eflux}).
Figure~\ref{fig:flux_tevol} shows time evolution of 
(a) mass flux,  
along with the areal star formation rate ($\Sigma_{\rm SFR}$),
and (b) energy flux.  
Every star formation burst is followed by a burst of energy
injection, and {this burstiness is reflected in large temporal
  variations in the mass and energy fluxes. The fluxes can become
  negative, meaning that the mass and/or energy of gas flowing inward
  exceeds that of the gas flowing outward, at a given height.} 
As distance $d$ from the midplane increases, the 
net mass flux significantly decreases, 
while the net energy flux even through $d=2$ and $3\kpc$ remains large.
While net negative mass fluxes (implying fallback) occur at $d=1\kpc$
after each burst of mass outflow,
net energy fluxes almost always remain positive.  The differing
behavior of mass and energy fluxes is a signature of multiphase flows. 

{Although star formation and hence SN feedback are impulsive 
  rather than continuous, the system
  approaches a quasi-equilibrium state.
  This state is a limit cycle mediated by the feedback loop, in which
  epochs of cooling and collapse alternate with epochs of heating and
  expansion.\footnote{{We note, however, that a quasi-steady
      state is not guaranteed for all galactic conditions 
      \citep[e.g.,][]{2017MNRAS.467.2301T}, and may
      only hold in parts of the parameter space in which the vertical
      oscillation time (which controls collapse and star formation) is
      sufficiently long compared to the stellar evolution timescale
      (which controls feedback and expansion).}}
Given that a
  quasi-steady state exists in the present simulation,
  horizontal- and temporal averages can be
  constructed to characterize this mean state.  Since different
  thermal phases coexist at all heights, to understand the outflows
  and inflows of mass and energy it is further necessary to separately
  construct horizontal averages of each thermal phase.}

\begin{figure}
\plotone{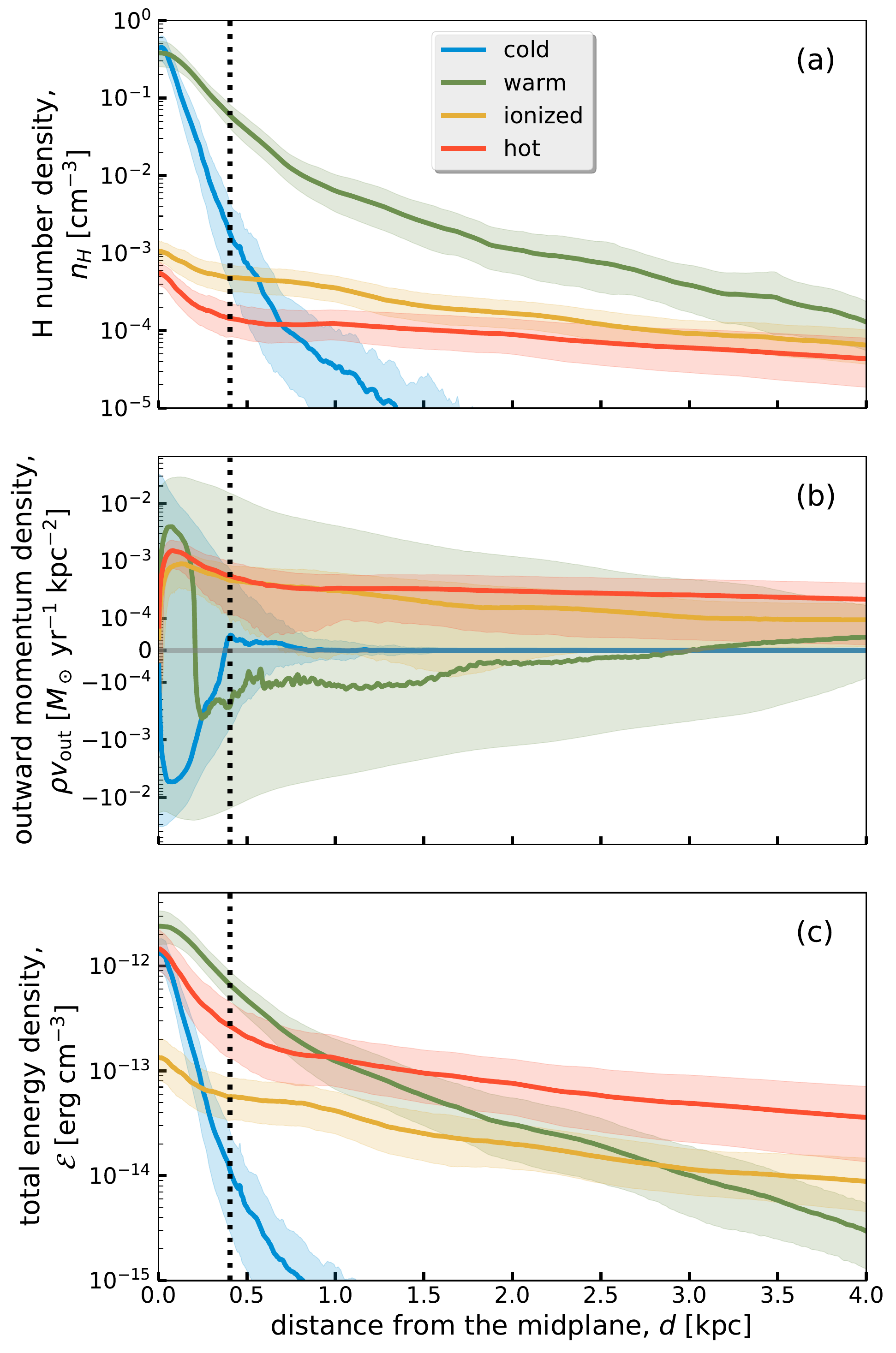}
\caption{Vertical distributions of 
  mass, vertical momentum, and energy densities, averaged horizontally
  and over upper ($z>0$) and
lower ($z<0$) sides of the disk
and over time $t=250-500\Myr$.
Colored lines for separate thermal components
show profiles of (a) hydrogen number density $n_H$,
(b) vertical momentum density
$\rho v_{\rm out}$ (which is the same as the mass flux), and
(c) total energy density (excluding gravity) $\cal E$, each as a function of 
distance $d$ from the midplane.  Color-coded shaded regions represent
one-sigma temporal fluctuations. In order to properly
visualize both the magnitude and sign of the momentum density,
we use a linear scale
for $|\rho \vout|<10^{-4}$ and a log scale for $|\rho\vout|>10^{-4}$.
The warm and hot phases respectively dominate mass and energy
densities above the disk scale height (indicated by vertical dotted line),
and the hot component also has the largest vertical momentum density
(net outward mass flux). 
\label{fig:zprof}}
\end{figure}

In Figure~\ref{fig:zprof}
we plot 
horizontally- and temporally-averaged 
profiles of
mass, momentum, and energy distributions for thermally separated gas phases;
{these profiles average over $t=250-500\Myr$ } and also average
over upper ($z>0$) and lower ($z<0$) sides.  Profiles show 
hydrogen number density, $n_H\equiv\rho/(\mu_Hm_H)$, 
outward vertical momentum density, $\rho \vout\equiv \rho v_z{\rm sign}(z)$, 
and total energy density (excluding gravity) 
\begin{equation}\label{eq:Etot}
  \mathcal{E}\equiv \frac{1}{2}\rho v^2 +\frac{P}{\gamma-1} + 
  \frac{B^2}{8\pi}.
\end{equation}
{Note that this energy density differs from the gas density multiplied by
  the Bernoulli parameter of Equation (\ref{eq:ber}),
as it includes
just thermal energy (rather than enthalpy) and also includes
 magnetic energy density.}

The four thermal phases plotted in Figure~\ref{fig:zprof} are 
cold ($T<5050\Kel$), warm ($5050\Kel<T<2\times10^4\Kel$), ionized 
($2\times10^4\Kel<T<5\times10^5\Kel$), and hot ($T>5\times10^5\Kel$).\footnote{
Note that we omit the ``unstable'' phase ($184\Kel < T < 5050\Kel$)
defined in Paper~I and merge it into the ``cold'' phase since (1) these phases
are not of primary interest in this paper since we are focusing on gas
above the disk scale height, and 
(2) the sum of these two phases numerically converges better than the individual
phases.}
Above the warm/cold layer ($d>H$, where $H\approx 400\pc$), 
the mass density is dominated by the warm component and the
energy density is dominated by the hot component.
As the individual terms in the energy density 
are proportional to corresponding terms in the momentum flux ($\rho v^2$, $P$),
the hot medium also
dominates the momentum flux
away from the midplane.  
The hot medium is the largest contributor to the
time-averaged vertical momentum
density, which is the same as the time-averaged net mass flux.
Although the mean value of vertical momentum density (or net mass flux)
of the warm medium is effectively
zero, there are
large temporal fluctuations (indicated by the green shaded region)
at small and intermediate $d$ 
because the warm gas contributes significantly to 
both outgoing and incoming mass fluxes at different times
(as in Figure~\ref{fig:flux_tevol}).  

The phase-separated momentum and energy density profiles in
Figure~\ref{fig:zprof} (and
corresponding profiles of mass and momentum flux) reflect essential
differences of gas flow dynamics between the warm and hot phases,
which we separately analyze in the following sections.

\section{Hot Winds}\label{sec:wind}

In this section, we focus on the hot component, defined by $T>5\times10^5\Kel$,
representing gas that has been shock heated by SN blastwaves.
The hot medium fills most of the volume above the disk scale height,
and cooling in tenuous hot gas is inefficient.  With pressure
gradients that accelerate it outward (cf. Figure~\ref{fig:zprof}(c)) and
a source from SN shocks propagating into the surrounding warm and cold medium
near the midplane, 
the {horizontally and temporally averaged} hot medium naturally fits the criteria for a (quasi-)steady, adiabatic wind.
We therefore might expect that the mass flux and energy flux
(and thus the Bernoulli parameter)
of the time-averaged hot component would be (approximately)
conserved as the gas flows outward.

\begin{figure*}
\plotone{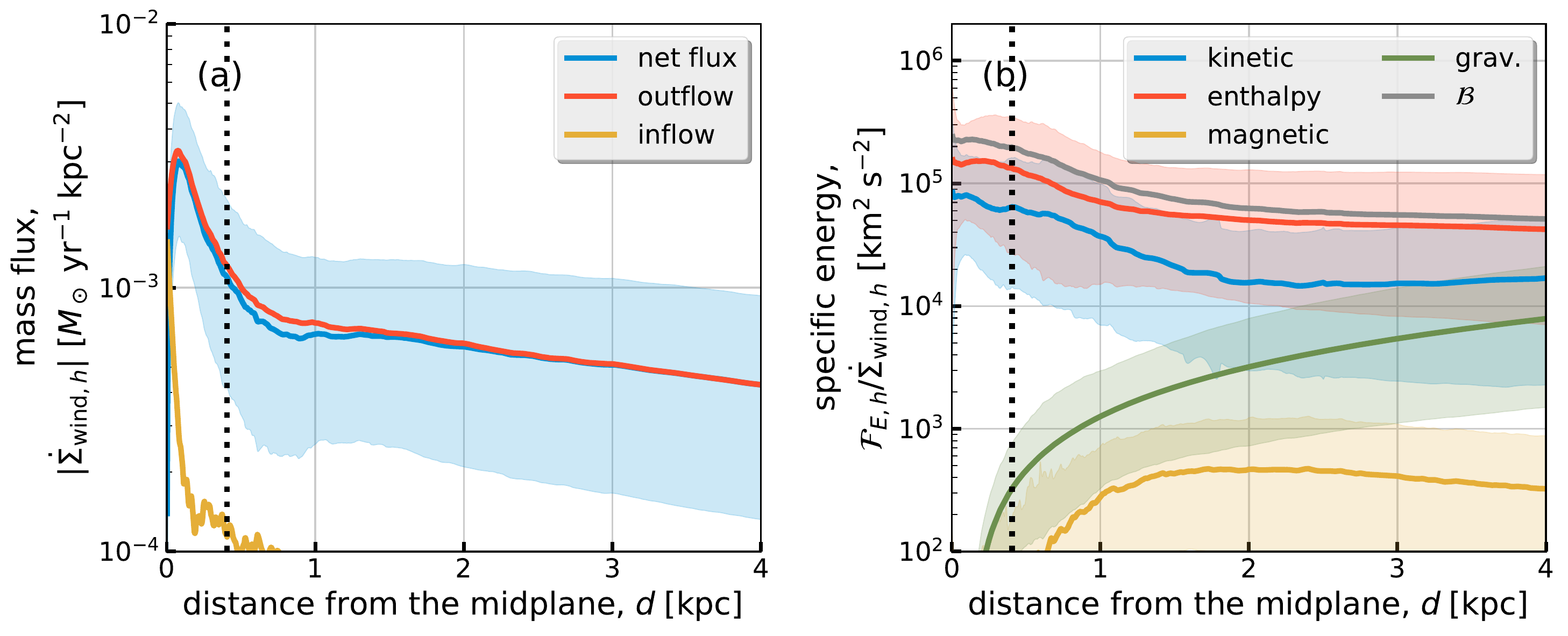}
\caption{Time averaged ($t=250-500\Myr$) vertical profiles 
of hot gas (a) mass fluxes and (b) specific energies.
(a) The net mass flux is shown as a blue line with
one-sigma temporal fluctuations as the blue
shaded region. The hot gas does not show any
significant inflows (yellow line), 
implying that any decrement of the mass flux is due to
a phase transition to cooler phases and not direct inflows of the hot gas. 
(b) Total Bernoulli parameter $\cal B$
as well as individual components
(i.e. kinetic, thermal, gravitational, and magnetic
terms; see Equations (\ref{eq:ber}) and (\ref{eq:spec_mag})).
Mean values are shown as colored lines,
and one-sigma temporal fluctuations are shown as shaded regions.
The specific enthalpy
and kinetic energy are much larger than the gravitational potential. 
The magnetic term plays a minor role. 
In both (a) and (b), the vertical dotted line indicates the disk scale
height ($H=400\pc$).
\label{fig:hot}}
\end{figure*}

Figure~\ref{fig:hot} explicitly shows vertical profiles
of (a) the mass flux, $\Mflux{,h}$
and (b) the specific energy, $\Eflux{,h}/\Mflux{,h}$,
based on time averages of each horizontally averaged flux.
{To distinguish between outflows and inflows, for the mass flux
  in addition to net flux we separately show the flux of outflowing
  ($sign(v_z)=sign(z)$) and inflowing   ($sign(v_z)=-sign(z)$) gas.}

\begin{figure*}
\plotone{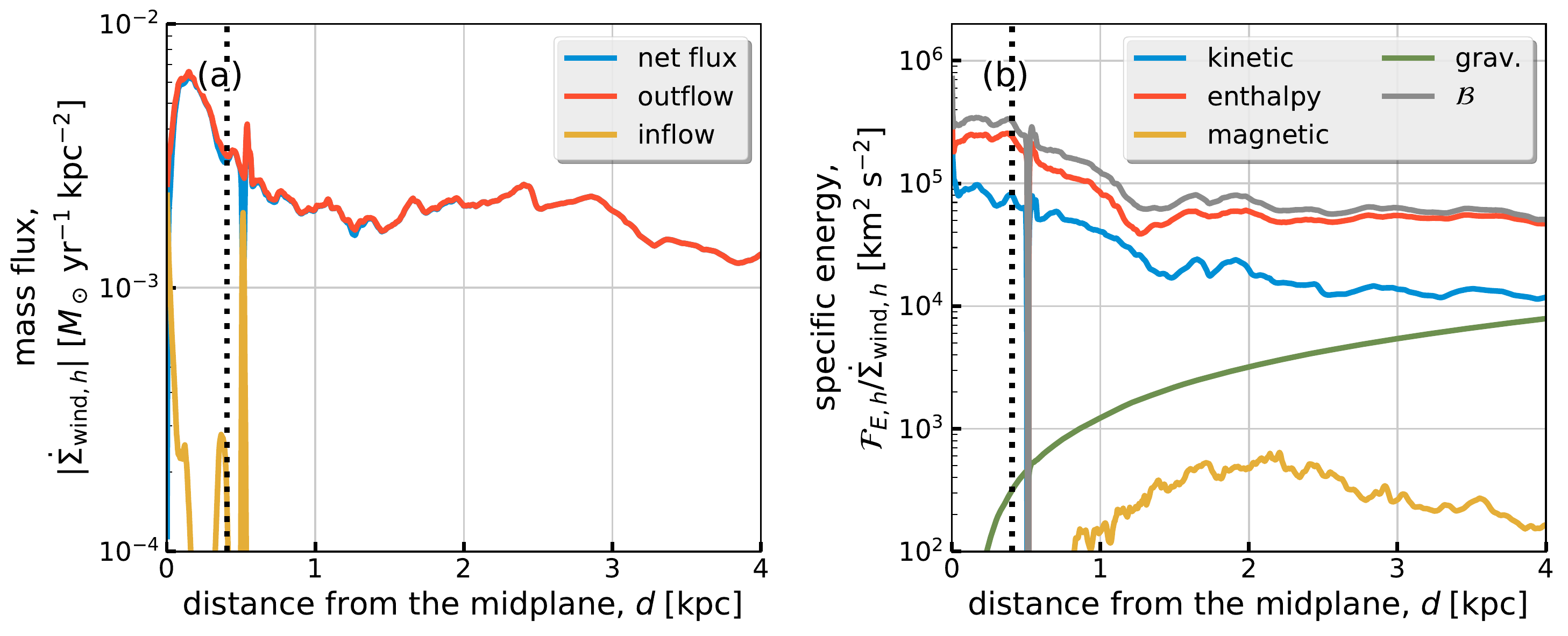}
\caption{
{  
  Mass flux and specific energy profiles for the hot wind as in
  Figure~\ref{fig:hot}, except at a single instant $t=300 \Myr$.}  
\label{fig:hot300}}
\end{figure*}

The mass flux of the hot gas shows net outflow
(blue line in Figure~\ref{fig:hot}(a)),
with negligible inflow flux at all heights 
(yellow line in Figure~\ref{fig:hot}(a)).
As the SNe that create the hot medium are concentrated near the midplane,
the hot gas mass flux within the warm/cold layer 
($d<H$; below the vertical dotted line in Figure~\ref{fig:hot})
first increases with $d$
and then decreases as
hot SNRs have a maximum size before the onset of cooling.
Above the point where the hot gas in the interior of SN remnants
(or superbubbles) breaks out of the warm/cold layer the hot gas
mass flux is nearly constant with $d$.
The slight decrease of the mass flux at large $d$ is caused by non-zero cooling
(radiative and mixing of hot gas with cooler phases).

Figure~\ref{fig:hot}(b) shows near constancy of the time-averaged
total specific energy $\mathcal{B}$ of the hot gas 
above $d=1\kpc$, as expected for a steady, adiabatic wind
(given the negligible Poynting flux).
We also calculate individual components of the (averaged) energy flux, 
and divide them by the (averaged) mass flux.
This provides the components of the mass-flux-weighted specific energy 
$\Eflux{,h}/\Mflux{,h}$, consisting
of kinetic, thermal (enthalpy), gravitational,  
and magnetic terms
(see Equations \ref{eq:ber} and \ref{eq:a_Poynting_z}).   
Once the hot gas breaks out of the warm/cold layer, the
gas flow approximately preserves mass and energy fluxes
because except for limited cooling 
there are no sources (or sinks) for the hot gas mass and energy.
We have checked that the individual cooling and heating terms, 
including Reynolds and Maxwell stresses arise from the shearing box
(see Equation 
(\ref{eq:a_energyz})), are indeed small compared to the SN energy injection. 

From $d=2-4\kpc$, the enthalpy of the hot gas implies a temperature in the
range $1.2-1.5\times10^6\Kel$ 
(or sound speed in the range $130-150\kms$), and the kinetic energy
of the hot gas translates to a velocity of $170-200\kms$.  The Poynting flux
contribution to $\mathcal{B}$ is negligible, corresponding to an Alfv\'en
speed of $30-35\kms$. 

In SN-driven hot winds, the enthalpy (specific heat) term 
dominates over other components
of the Bernoulli parameter, including the gravitational potential, at
heights less than the disk radius.
At large distance where streamlines open up in angle,  
one could expect hot galactic winds
with constant $\cal B$ to accelerate past a sonic point
to an asymptotic velocity of
$v_{\rm wind}\equiv\sqrt{2}(\mathcal{B}-\Phi)^{1/2}$ as they adiabatically cool,
similar to classical Parker stellar wind solutions.\footnote{Note that for galactic potentials, $\Phi$ is generally
  computed relative to the midplane, whereas for point mass potentials
  the $\Phi=0$ reference point is at infinite distance; the zero point in
  $\Phi$ is irrelevant as long as it is consistent in $v_{\rm wind}$ and
  $\mathcal{B}$.}
{Global simulations indeed show the expected behavior at large distance
\citep[e.g.][]{2017MNRAS.470L..39F}.}

In general, hot
winds are accelerated by pressure gradients at the same time as
enthalpy is converted to kinetic energy, and a sonic transition in
a steady wind is only possible if the crossectional area increases as
the flow moves outward \citep[e.g.][]{1992pavi.book.....S}.
In the Cartesian geometry of the present simulations, streamlines cannot
open up and there is no associated adiabatic cooling, limiting the pressure
and density gradients and therefore the acceleration of the  flow.
However, the constancy of both the Bernoulli parameter and mass flux with
$d$ for the hot medium in our simulation suggests
that it properly represents the near-disk regions for a generalized galactic
disk wind,
in which streamlines emerge from the disk vertically (with $A=const$
when $z\ll R_{\rm disk}$) and would open up ($A$ increasing with distance) when
$z\simgt R_{\rm disk}$.
{The mass and energy fluxes carried by the 
hot wind are controlled} by the
interaction between SN shocks and the warm-cold medium
that creates the hot ISM well within the disk, in processes 
that are unlikely to be affected by large-scale global galactic and
CGM properties and geometry.  Therefore, the
Bernoulli parameter we calculate
should be a robust estimator of asymptotic {hot} wind speed 
irrespective of the constraints of our Cartesian box.

As shown in Figure~\ref{fig:flux_tevol} 
(and demonstrated by the shaded region in Figure~\ref{fig:hot}(b)), 
star formation and hence outflows are bursty, 
resulting in large temporal fluctuations.
Based on analysis of 
{${\cal B}$ for the hot component  within}
short time ranges,
the maximum asymptotic velocity\footnote{Here, this maximum wind velocity
is calculated using $\Phi$ at a height $d=4\kpc$, although in reality the wind 
velocity would continue to decrease slowly with distance due to
the logarithmic increase of $\Phi(r)$
with $r$ at large distances in dark matter halo potentials.}
of the hot wind could
reach up to $v_{\rm wind}\sim 500\kms$, while the
mean asymptotic wind speed would be $v_{\rm wind}\sim 350\kms$.
Defining the speed required to escape to $\bf r$ as 
$v_{\rm esc}({\bf r}) \equiv [2(\Phi({\bf r})-\Phi(0))]^{1/2}$,
in our simulation domain
$v_{\rm esc}(z=4\kpc)\sim 130\kms$, so the
hot wind easily escapes, 
{i.e. ${\cal B}\gg \Phi$ even at large $d$} for the hot component.  
A hot wind launched with the local conditions of our simulation would
also be able to propagate far into the halo for the Milky Way,
where the escape velocities are $v_{\rm esc}(50\kpc)\sim 350\kms$ and
$v_{\rm esc}(150\kpc)\sim450\kms$ 
using {\tt MWPotential2014} in {\tt galpy} \citep{2015ApJS..216...29B}.
More generally, the far-field velocity for a hot wind with given local
launching conditions can be estimated based on $\mathcal{B}$
and the large-scale galactic potential.  

{Finally, we note that while we have mostly based the discussion above
on temporal averages of horizontally-averaged profiles, the properties
of instantaneous profiles are quite similar.  To illustrate this,
Figure \ref{fig:hot300} shows the mass flux and specific energy of the
hot component at $t=300$ Myr, the ``outflow'' snapshot shown in Fig.
(\ref{fig:slice}).  Except for local fluctuations, the instantaneous
mass flux and specific energy profiles are overall very similar to the
corresponding time-averaged profiles.  In particular, the specific energy
profiles in Fig. \ref{fig:hot} and \ref{fig:hot300} are quantitatively
almost the same, while the mass flux is higher in Fig. \ref{fig:hot300} than in
Fig. \ref{fig:hot} because the latter includes non-outburst epochs as well as
outburst epochs in the temporal averaging.}

\section{Warm Fountains}\label{sec:fountain}

Over the duration of the simulation, 
the mean net mass flux in warm gas out of the simulation domain is
$\sim 1.1\times10^{-4} \sfrunit$, about 28 percent of the net mass flux in hot gas.
As we shall show, the outward
mass flux of the warm medium secularly decreases
with $d$, such that if our box were taller we would expect the mean net
mass flux out of the simulation domain to be even smaller.
{With negligible time-averaged net outward mass flux at large distance,
  the warm medium at $d\simgt 1\kpc$ in our simulation is not a true galactic wind.}
Even though the inflows and outflows of warm gas over long timescales
are essentially balanced, time variations in the warm gas flux
$\rho v_{\rm out}$ are
quite large and include both positive and negative values
(green shaded region of Figure~\ref{fig:zprof}(b)).
{The fluctuating behavior of the warm medium
  can be contrasted with the much smaller temporal fluctuations of the
  mass flux in the hot gas (red shaded region of Figure~\ref{fig:zprof}(b)).}
Evidently, the warm gas does not produce an escaping
wind like that in the hot gas
but a fluctuating fountain that at any time consists of both outflow and inflow.
{Figure~\ref{fig:flux_tevol} shows that alternating inflow
  and outflow dominance in the warm gas is reflected in the alternating signs
  of the mass flux for the whole medium at $d=1\kpc$.  In addition,
  the decrease in the magnitude of the total (phase-integrated) 
  mass flux with increasing
  $d$ reflects the secular decrease in the net mass flux of the warm fountain
  (difference between inward and outward fluxes) with distance.}

\begin{figure*}
\plotone{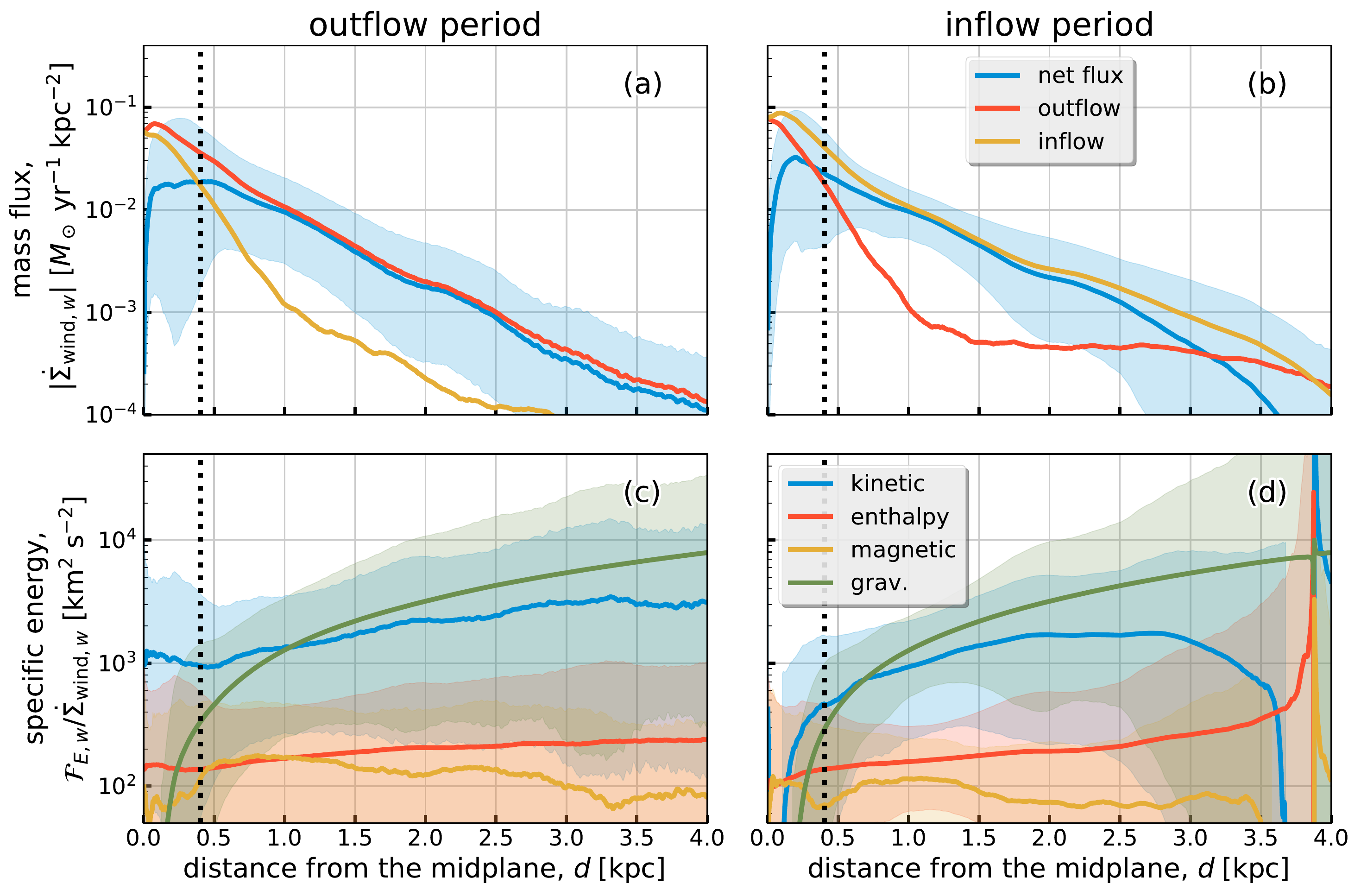}
\caption{Vertical profiles of warm gas mass fluxes (top)
  and specific energies (bottom) 
averaged
over outflow (left; $\Mflux{}(1\kpc)>0.001\sfrunit$) and inflow 
(right; $\Mflux{}(1\kpc)<-0.001\sfrunit$) periods.
Inflows and outflows are comparable with each other,
implying little net mass and energy outflows (or inflows)
from the simulation domain associated with the warm medium. 
Note that, for visualization purpose, we take absolute values 
of the inflow mass flux and net flux during the inflow period;
these have negative signs by definition.
For the warm medium, the kinetic term exceeds both the thermal and
magnetic terms in the specific energy, but is always lower than the
gravitational term above $\sim 1\kpc$.  This explains why the
warm medium creates a fountain rather than a wind.  
The vertical dotted lines in all panels indicates the gas scale
height ($H=400\pc$).
\label{fig:warm_flux}}
\end{figure*}

To quantify the characteristics of warm fountain flows, 
we take time averages selectively for 
outflow ($\Mflux{}(1\kpc)>0.001\sfrunit$)
and inflow ($\Mflux{}(1\kpc)<-0.001\sfrunit$) periods.
Outflows of warm gas occur when many correlated SNe from a 
star formation burst lead to a  
superbubble expanding into the warm and cold layer, while 
inflows occur when the disk is in a quiescent state with reduced
star formation after the cold medium has been dispersed by a
previous burst.
Figure~\ref{fig:warm_flux} plots time averaged mass flux (top) and 
specific energy (bottom) for outflow (left) and inflow (right) periods. 
Although one flow dominates the other during each period, 
the opposite flows always exist at all heights 
and are more significant compared to the case for
hot gas (Figure~\ref{fig:hot}).
For the warm medium, the kinetic term in the specific energy
exceeds the enthalpy,  
but remains below the gravitational potential term at $d\simgt 1\kpc$.  
This explains why most of
the warm gas outflow turns around at $d\sim1-2\kpc$ and
falls back toward the midplane.
During the outflow period, the mean velocity of the warm gas is in the
range $\sim60-80\kms$ for $d=1-4\kpc$.

Overall, the warm medium occupies more volume above $d=1\kpc$ 
during the inflow period than the outflow period.
This is because the hot gas is mainly generated
during the outflow period by shocking the warm gas, and
the high-pressure hot gas confines the warm gas into small cloudlets.
When the disk becomes quiescent, the warm gas expands into previous
hot wind channels
(see Figure~\ref{fig:slice}).
About a factor of 5 to 10 more volume
is occupied
by the warm gas in the inflow period than the outflow period for 
$d=2$ and $3\kpc$ slabs.

\begin{figure*}
\plotone{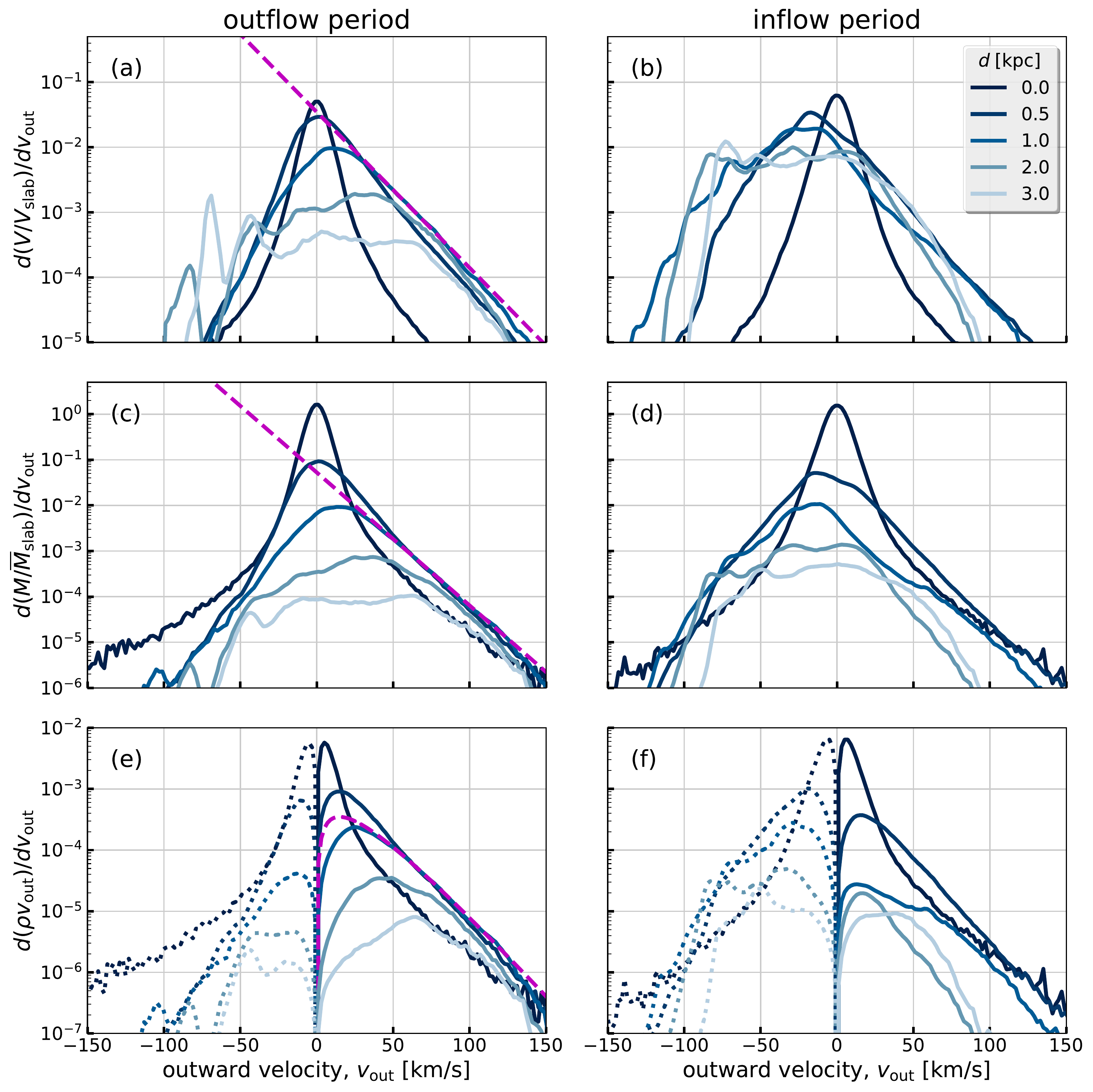}
\caption{Outward vertical velocity ($\vout\equiv v_z{\rm sign}(z)$) 
probability distributions of the warm medium.
Distributions of volume (top), 
mass (middle), and mass flux (bottom) 
as a function of velocity 
are calculated for $20\pc$ thickness slabs above and below the
midplane ($z=\pm d$) {centered at $d=0,0.5,1,2,3\kpc$.}
In the bottom row, the mass flux within each velocity bin is in units of $\sfrunit/\kms$, and 
negative mass fluxes are shown with dotted lines.
Warm gas with both signs of $\vout$ is present
during both ``outflow'' and ``inflow'' periods, but the corresponding
mean velocity {(for $d\ne 0\kpc$)} changes sign.  
Acceleration of the warm gas to high velocities,
especially during outflow periods, is evident in the difference of 
profile shapes between the midplane ($d=0\kpc$) and higher
latitude, with an exponential tail at high velocity developing by $d=1\kpc$. 
However,
an overall deceleration with height in the increasing gravitational potential
is also evident in comparison of profiles at increasing $d\ge 1\kpc$.
The magenta dashed lines in the left column 
are fits to the fast-moving gas $\vout>50\kms$ at $d=1\kpc$
given by {Equations (\ref{eq:fit}) and (\ref{eq:fit_mf})}.
\label{fig:warm_vpdf}}
\end{figure*}

Although in the current simulations the warm gas is
almost entirely 
confined by the 
galactic gravitational potential,
this would not necessarily be true if the potential were shallower,
as in dwarf galaxies.
Direct simulations for different galactic conditions, including
shallower potentials, are underway using the same TIGRESS framework.
However, we can also use our 
current simulation to provide information on what might be expected
by quantifying the fraction of fast-moving warm gas.
Figure~\ref{fig:warm_vpdf}
shows outward velocity probability distribution functions (PDFs)
weighted by volume (top),
mass (middle), and mass flux (bottom)
within slabs of thickness $\Delta_{\rm slab}=20\pc$ 
during outflow (left) and inflow (right) periods.
We show results at several distances $d$ averaged over
both sides of the disk ($z=\pm d$).
Note that the volume-weighted PDF is normalized by slab volume 
($V_{\rm slab}=L_xL_y2\Delta_{\rm slab}$),
and the mass-weighted PDF is normalized by the mean mass within the same
volume 
($\overline{M}_{\rm slab}\equiv M_{\rm tot} 2\Delta_{\rm slab}/L_z$)
($M_{\rm tot}$ is the mass in the whole domain at a given time),
while the mass flux PDF is in physical units of $\sfrunit/\kms$.
During both outflow and inflow periods,
Figure~\ref{fig:warm_vpdf} shows that the warm gas velocity
has a broad distribution with both outward and inward velocities.

During the outflow period, Figures~\ref{fig:warm_vpdf}(a) and (c) show that 
the volume and mass of high velocity ($>50\kms$) warm gas increases 
from the midplane to $1\kpc$, where the specific kinetic
energy is larger than the gravitational potential
(see Figure~\ref{fig:warm_flux}(c)).
The increase in mass of high-velocity, high-altitude warm gas 
between the midplane and $d=1\kpc$ 
is due to acceleration of the warm medium pushed by expanding
superbubbles; this includes warm gas that was shock-heated to the hot phase
(and accelerated to high velocity) and subsequently cooled back down.
Figures~\ref{fig:warm_vpdf}(c) and (e) show that as $d$ increases 
the peaks of the distributions of mass and mass flux move to higher velocity
and the overall outflowing gas fraction decreases.
This general trend represents
dropout/turnaround of warm gas fluid elements with low (and
decelerating) velocities that are unable to climb to large $d$ in the 
gravitational potential.

In principle, acceleration of warm clouds driven by hot-gas ram pressure,
cooling of fast hot gas, and dropout of low-velocity warm fluid elements
could  all contribute to the gradual increase of warm-medium
specific kinetic energy at $d>H$ shown in Figure~\ref{fig:warm_flux}(c).
Figure~\ref{fig:warm_vpdf}(c) shows, however, that overall the mass of
high-velocity warm gas is {\it decreasing} with increasing $d$.  This has
the important implications that in our simulation (1)
warm clouds at large $d$ are {\it not} significantly
accelerated by ram pressure of the hot, high-velocity gas that
is flowing out around them, and
(2) relatively little hot gas is converted to the warm phase through
cooling at large $d$.  Rather, the warm medium is primarily
accelerated via direct energy input from SNe at $d\simlt 1\kpc$, and at higher
altitudes warm fluid elements slow and turn around according to the
competition between the gravitational potential and the kinetic energy they
initially acquired at small $d$.
Figure~\ref{fig:zprof}(c) and Figure~\ref{fig:hot}(b) are also telling in this
regard:  the total energy density (and individual components)
of the hot medium declines very slowly for $d\simgt 1\kpc$,
while the energy density
of the warm medium declines steeply; since momentum flux {terms} are
proportional to energy density {terms},
this indicates that there is no significant transfer of
momentum from the hot to the warm gas.

During the inflow period, the majority of the warm gas is falling.
Since SN feedback is never completely turned off, however,
some warm gas is still accelerated outward. 
The fraction of (outgoing) fast-moving warm gas ($v_{\rm out} > 50\kms$)
is reduced by a factor of 5 to 10
in inflow compared to outflow periods.  
Combined with the total outflowing mass fluxes of 
$\Mflux{,w}\sim 10^{-2}$ and $10^{-3}\sfrunit$
 at $d=1\kpc$ during outflow and inflow periods
(see Figure~\ref{fig:warm_flux}(a,b)), respectively, 
the mass fluxes of the fast-moving warm gas are about $6\times10^{-4}$
and $10^{-5}\sfrunit$
(see Figure~\ref{fig:warm_vpdf}(e) and (f)).  
This can be compared to a mean mass flux of
hot gas at the same height of
$\sim 1.3\times10^{-3}$ and $2\times10^{-4}\sfrunit$ during outflow
and inflow periods, respectively.
Although the ``fast'' warm outflow has comparable mass flux
at $d=1\kpc$ to that of the hot
medium during outflow periods,
even at $v_{\rm out} > 50\kms$ the warm medium yields 
very little mass escaping from the large-scale potential in the present
simulation, while the hot medium mostly escapes
($0.12$ and $0.034\Surf$ of hot and warm gas have respectively escaped 
during the time interval of $250\Myr$).  This emphasizes
the importance of measuring not just mass fluxes, but mass fluxes and
specific energies in comparison to galactic escape speeds.  

At large distances the large-scale gravitational potential
strongly affects the warm-gas velocity distribution by enforcing
dropout of lower-velocity material, but closer to the midplane
this is less of an issue.  At around the disk scale height, the 
gravitational potential is small compared to the specific kinetic energy
of the gas, and global geometric effects are not important yet.
We therefore consider the velocity distribution at $d=1\kpc$ as
representative of the launching conditions for a warm outflow,
which would apply relatively independently of the global galaxy (e.g.
in a dwarf as well as a large galaxy for given local conditions). 
Here, we find the PDFs during outflow periods of the fast-moving warm gas
($\vout > 50\kms$) 
at $d=1\kpc$ are well fitted by a single exponential function,
\begin{equation}\label{eq:fit}
\frac{df}{dv_{\rm out}} = A_f\exp\rbrackets{-\frac{v_{\rm out}}{v_f}}, 
\end{equation}
{for $f=V/V_{\rm slab}$ or $M/\overline{M}_{\rm slab}$,
where the normalization factors for volume and mass weighted PDFs are
$A_V= 0.63/v_V$ and $A_M=0.79/v_M$, respectively, and the characteristic velocities are
$v_V=18\kms$ and $v_M=15\kms$.
Using the mass PDFs, the mass-flux PDF is given by
\begin{equation}\label{eq:fit_mf}
\frac{d(\rho v_{\rm out})}{dv_{\rm out}}=
\frac{\overline{M}_{\rm slab}}{V_{\rm slab}} A_M v_{\rm out} \exp\rbrackets{-\frac{v_{\rm out}}{v_M}},
\end{equation}
where the mean density, $\overline{M}_{\rm slab}/V_{\rm slab}$,
is given by $\rho = 1.2 \times10^{-3} \Msun \pc^{-3}$, 
corresponding to hydrogen number density $n_H=0.033\pcc$.}

By quantifying the mass flux PDF in 
warm gas for models with different local conditions, it should be
possible to develop a comprehensive quantitative characterization of
warm wind launching by star formation feedback. 
{These local results
could then be used to make global predictions.
  For example, integration of Equation (\ref{eq:fit_mf}) for velocities
  $\vout > v_H$ for would yield a mass flux
  $(v_H/v_M + 1) \exp(-v_H/v_M) \dot\Sigma_{\rm wind,w}(d=1\kpc)$.
If this holds in general, it means that
  measurement of the launching-region warm mass flux
  $\dot\Sigma_{\rm wind,w}(d=1\kpc)$ for given local disk conditions could be used to
  predict the flux that actually escapes into the halo for a galaxy with
  arbitrary halo velocity $v_H$.
  }
This statistical
characterization can be combined with measurements of mass flux and
Bernoulli parameter for the hot medium to develop subgrid models
of multiphase wind driving for implementation in galaxy formation simulations.

\section{Mass and Energy Loading}\label{sec:loading}

\subsection{Simulation Results}\label{sec:loading_result}

In this section, we discuss key quantities of 
gas outflows driven by SNe, mass and energy loading factors.
The mass loading factor, $\beta$, through the surfaces at $d$ 
(including both sides of the disk plane)
is conventionally defined by the ratio of ``outgoing''
mass flux to the star formation rate as
\begin{equation}\label{eq:mass_loading}
\beta\equiv \frac{\Mflux{}^+}{\Sigma_{\rm SFR}},
\end{equation}
while the energy loading factor, $\alpha$, is defined by the ratio of ``outgoing''
kinetic + thermal (enthalpy) energy flux to the energy production rate of SNe,
\begin{equation}\label{eq:energy_loading}
\alpha\equiv \frac{\Eflux[KE]{}^++\Eflux[TE]{}^+}{E_{\rm SN}\Sigma_{\rm SFR}/m_*},
\end{equation}
where $E_{\rm SN}=10^{51}\erg$ is the total energy per SN, and 
$m_*=95.5\Msun$ is the total mass of new stars per SN (see Paper~I).
To compute $\Mflux{}^+$, $\Eflux[KE]{}^+$, and $\Eflux[TE]{}^+$ we
select only zones with outflowing gas, 
i.e. with ${\rm sign}(v_z)={\rm sign}(z)$.  
The areal star formation rate averaged over $t=250-500\Myr$ is $\Sigma_{\rm SFR}=0.006\sfrunit$;
we use this average $\Sigma_{\rm SFR}$ in computing all loading factors.  
Note that a rolling mean of $\Sigma_{\rm SFR}$ with 
100Myr time window gives 50\% variation with respect to the mean.
Because there is generally a temporal offset between star formation bursts
and winds (see Figure~\ref{fig:flux_tevol}),
the instantaneous ratio of mass or energy fluxes to the star
formation rate can significantly over- or under-estimate the true physical
loading.  The ratio of time-averaged wind fluxes to the time-averaged star
formation rate is more meaningful.   

\begin{figure*}
\plotone{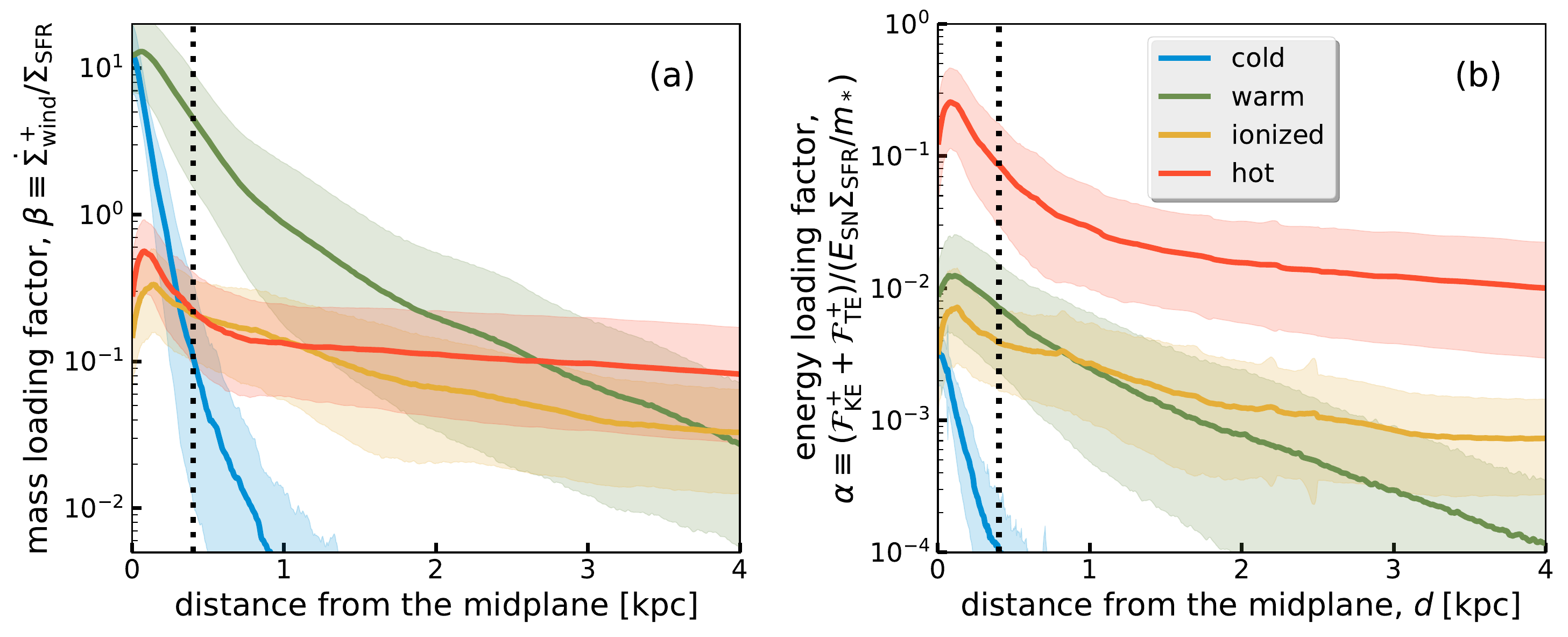}
\caption{Mass and energy loading factors in each thermal phase.
For both cold and warm gas, 
both mass and energy loading factors stiffly decrease as a function of $d$.
Mass loading also drops off with $d$ for the ionized phase.  
Only the hot gas is a true wind, with well-defined mass and
energy loading factors of $\beta_h\sim0.1$ and $\alpha_h\sim0.02$, respectively.
\label{fig:loading}}
\end{figure*}

We decompose the gas into four thermal components and present the loading factors
of each thermal phase as a function of $d$ (Figure~\ref{fig:loading}).
The energy flux is always dominated by the hot gas, 
with the energy loading factor of $\alpha_h\sim0.01-0.05$ above $d>1\kpc$.
SN feedback causes large outgoing mass fluxes of warm and cold gas 
within the disk scale height $d<H$,
but the majority of the warm and cold medium has low velocity and
cannot travel far from the midplane; this mass flux at small $d$
is best thought of
as the ``upwelling'' of turbulent motions within the disk
driven by expanding SNRs and
superbubbles.\footnote{Allowing for the work done by shock-heated hot gas,
both isolated and spatially correlated SNe
  inject a mean spherical momentum/SN to the ISM of
  $\sim 10^5 \Msun\kms$ \citep[see][and references therein]{2015ApJ...802...99K,2017ApJ...834...25K},
  an order of magnitude greater than the momenta
  of the initial SN ejecta.  Most of this momentum goes into maintaining
  quasi-equilibrium force balance with gravity
  in the bulk of the ISM, rather than driving a wind.}
The mass loading factor of the ionized gas also decreases significantly
with $d$.
By the time the flow reaches the edge of the simulation domain,
$\beta_h$ is at least a factor $\sim3$
larger than all the other components.   Therefore, 
as we have concluded in Sections~\ref{sec:wind} and \ref{sec:fountain}, 
only the hot gas forms a genuine``galactic wind,''
with a mass loading factor of 
$\beta_h\sim 0.1$ that is nearly constant above $d>1\kpc$.
{We note that with the above definitions of $\alpha$ and $\beta$,
  the Bernoulli parameter for the hot wind is given by
  ${\cal B} = \Phi + (\alpha_h/\beta_h)E_{\rm SN}/m_*$}.

The vertical dependence of the mass loading factors
of non-hot phases shown in Figure~\ref{fig:loading}(a) 
implies that it is important to provide careful distinctions when
reporting on the outflows measured in numerical simulations, as
``wind'' mass loading can be greatly overestimated if this is not done.
In particular, the steep decrease of $\beta_w$ with $d$ implies that
if this were measured at $d\sim 1\kpc$ it would overestimate the value at
the edge of our box ($d \sim 4\kpc$) by a factor $\sim 30$, and the
true value at larger distance would be even smaller
\citep[see also][who reached similar conclusions about total $\beta$]{2016MNRAS.459.2311M}.
Thus, a measurement
of $\beta_w$ from a simulation with small vertical domain cannot by
itself provide a prediction for warm gas mass loss in a galactic wind.
However, even with a limited vertical domain, it is possible to discriminate
between fountain flow and wind by combining measurements of the warm
gas mass flux and its vertical velocity (as in
Section~\ref{sec:fountain}).  Similar considerations apply to observations of
gas at $T\sim 10^4\Kel$ at high latitudes in edge-on galaxies, but in that
case uncertain projection effects make this even more problematic:
without an unambiguous measurement of velocity (which is subject to assumed
wind geometry), it is impossible to distinguish between fountain
flow and wind from observations of the emission measure.

More generally, it is essential to decompose
outflows in simulations into separate thermal phases to distinguish
winds from fountain flows.  The integrated $\beta$ cannot be taken as
a true wind mass loading unless the measurement is made at very large
distance in a global simulation.  Nevertheless, individual measurements of 
mass fluxes in different phases, when combined with analysis of the
components of their individual Bernoulli parameters, can be used to
distinguish fountains from winds even within a limited vertical domain.

The mass loading factor of the hot gas obtained here 
is consistent with simple estimates based on idealized experiments 
of superbubbles driven by multiple SNe in the warm-cold ISM
\citep{2017ApJ...834...25K}.
The shock from an individual SN sweeps up $\sim 10^3\Msun$
before it cools and forms a shell \citep{2015ApJ...802...99K}.  For
superbubbles created by multiple SNe, the maximum mass in hot gas per SN
is $\hat M_h\sim 400-2000\Msun$ prior to shell formation, but subsequently 
this drops to $\hat M_h\sim 10-100\Msun$ (lower at higher ambient
density).  When star formation rates are self-regulated 
\citep[e.g.,][]{2013ApJ...776....1K,2015ApJ...815...67K},
the mean interval between SNe within projected area $\pi H^2$
in the disk of a galaxy is always $\sim 0.3\Myr$,
and breakout of superbubbles is expected to occur after shell formation
(see Section~5 of \citealt{2017ApJ...834...25K}).
For a SN interval $\ge 0.1\Myr$, Figure 11 of \citet{2017ApJ...834...25K}
shows that by the time superbubbles reach a radius of $2H$,
$\hat M_h \sim 10- 30 \Msun$, depending only weakly on ambient density. 
This corresponds to 
$\beta_h=\hat{M}_h/m_*\sim 0.1-0.3$.
The same idealized superbubble simulations show 
a hot-gas energy loading factor of a few percent when superbubbles
expand beyond $\sim H$, since most of energy has already been transferred
to acceleration and heating of ambient gas and lost via radiative cooling
at the time of shell formation.  

\citet{2017ApJ...834...25K} argued that 
$\beta_h< 1$  is expected quite generally 
unless the temporal and spatial correlations of SNe 
are extremely enhanced compared to their mean values,
requiring more than a factor of 40 elevation
compared to the average conditions in self-regulated galactic disks
where $\Sigma_{\rm SFR} \ge 10^{-3}\sfrunit$.
Although our simulation does exhibit large temporal fluctuations  
(see Figure~\ref{fig:flux_tevol}),
the peak upward fluctuation compared to the mean star formation rate
is only a factor of 5. Further systematic investigations for different
galactic conditions will be needed to confirm whether the
predictions for low $\beta_h$ apply universally in star-forming galaxies.
If local or global conditions make star formation inherently extremely
bursty, then $\beta_h$ may be higher.  

\subsection{Comparison with Observations}\label{sec:loading_obs}

The low mass loading factor of the hot gas in our simulations
is comparable to
({or slightly smaller than})
the observed mass loading factor estimated 
in the best studied local starburst M82. Using Chandra X-ray observations,
\citet{2009ApJ...697.2030S} constrained
the ``central'' mass loading factor of the hot gas to
about $\beta_h \sim 0.3-1$,
with the ``central'' energy loading factor about
$\alpha_h\sim0.3-1$.
Here, they constrained quantities using 
a large number of hydrodynamical models to explain 
the observed diffuse and hard X-rays, which come from the central 500pc region.
\footnote{However, we note that in their modeling the hot wind
  freely expands into a very tenuous medium rather than expanding
  into a dense ISM.  By comparison, it is evident
  e.g. in Fig. 8 of \citet{2017ApJ...834...25K} that before superbubble
  breakout from the warm/cold ISM, the hot gas has very high velocity but
  mostly remains subsonic. This suggests that lack of a dense surrounding
  medium in a hydrodynamic wind comparison model might lead to a
  Mach number higher than would be realistic, and hence
for density and temperature constrained by emission properties
  the mass loss rate could be overestimated by a factor of a few.}
It is difficult to make a direct comparison,
but if we consider the state of the hot medium within the 
energy-injecting layer ($d<H$) in our simulations, we have $\beta_h\sim 0.5$
and $\alpha_h\sim 0.1-0.2$. Although there are no systematic
observational studies of mass loading factors of the hot gas, 
$\beta_h<1$ is suggested for a wide range of star formation rates
from dwarf starbursts to ultraluminous infrared galaxies,
utilizing the \citet{1985Natur.317...44C} wind model 
with observational constraints of
the X-ray luminosity and star formation rates
\citep{2014ApJ...784...93Z}.

For the solar neighborhood conditions investigated in the present simulation,
the warm gas accelerated by the SNe cannot reach velocities
fast enough to escape the gravitational potential.
Typically, star formation bursts launch 
warm outflows with velocity up to $\sim 100\kms$,
but with most gas at lower velocity (see Figure~\ref{fig:warm_vpdf})
which can climb to $\sim 1\kpc$ but
no further (compare the kinetic and gravitational curves in
Figure~\ref{fig:warm_flux}(c)).
Without additional energy and momentum input at high-$|z|$, SN feedback alone
cannot drive warm winds in the Milky Way-type galaxies.
Above $d=1\kpc$, Figure~\ref{fig:loading}(a) shows that $\beta_w<1$.
Even though relatively fast warm gas could escape from dwarf galaxies
with a shallower potential well than the Milky Way,
Equation (\ref{eq:fit}) implies that $\beta_w\sim 1/3$
for $\vout>50\kms$ at most.   
This suggests that achieving $\beta>1$ in dwarfs, as appears necessary
to explain current-day galaxy-halo relationships and cosmic history,
would require an additional acceleration mechanism such as interaction
with an outflowing cosmic ray fluid 
\citep[e.g.,][]{2013ApJ...777L..38H,2016ApJ...816L..19G,2017ApJ...834..208R}.
In cosmic-ray driven winds, 
SN feedback may still be crucial for pushing warm gas out to
large $d$ where cosmic ray pressure gradients are sufficient to
produce efficient acceleration (Mao \& Ostriker 2017, submitted).

Observational studies of warm phase outflows 
(possibly including the ionized phase according to our definition)
indicate a wide range of the mass loading factor, $\beta_{w+i}\sim 0.1-10$
\citep[e.g.,][]{2015ApJ...809..147H,2017MNRAS.469.4831C},
with systematically decreasing
trends for increasing galaxy mass, circular velocity, and total SFR.
As the level of wind mass loading may vary with both local conditions
(including $\Sigma_{\rm gas}$ and $\Sigma_{\rm *}$) as well as the
global gravitational potential, it will be quite interesting to measure outflow
properties in simulations under widely varying galactic conditions,
for comparison with current empirical scaling trends and future observations.  

\subsection{Wind Driving Simulation Context}\label{sec:loading_sim}

Recently, a number of other research
groups have performed simulations with similar 
local Cartesian box setups to study galactic outflows driven by SNe 
\citep[e.g.,][]{2013MNRAS.429.1922C,2015MNRAS.446.2125C,2016ApJ...816L..19G,2016MNRAS.456.3432G,2016MNRAS.459.2311M,2017ApJ...841..101L,2017MNRAS.466.1903G}.
Most of these simulations have adopted fixed SN rates, 
while varying the SN placement (e.g. random vs. in high density regions).
In contrast, in our simulation, SN rates and locations are self-consistent with
star formation, which we believe is crucial in creating realistic
multiphase outflows.

\citet{2016MNRAS.456.3432G} ran a set of simulations with solar neighborhood
conditions, focusing on the effect of the SN placement and of SN clustering. 
The authors
emphasize that due to the short duration of their simulations ($100\Myr$),
definitive
conclusions cannot be drawn regarding wind driving.  However, their
measurements of outflow in the $d=1\kpc$ plane can be compared to
our fountain flow measurements.
For a range of different SN feedback treatments,
they report mass loading factors $\sim 5-10$.
Although this is larger than what we find
{($\beta_w\sim 1$ at $d\sim 1\kpc$)},
their simulations do show a decline in outflow over time,
{and our measurements are at $t>250\Myr$.}
As pointed out by the authors, the majority of the outflowing mass is
relatively dense
($n_H\sim 0.1$)
and moves slowly ($\vout\sim20-40\kms$),
similar to the properties 
of our warm fountain at low $d$. Their fraction of high-velocity
gas ($\vout>500\kms$) is only $2-8\times10^{-5}$, which correspondingly
reduces the mass loading factor of gas that is certain to escape to
large distance.

In simulations including cosmic rays,  
\citet{2016ApJ...816L..19G} found slowly moving 
($\vout\sim10-50\kms$) warm outflows with a mass loading factor near unity.
The SN-only comparison model of \citet{2016ApJ...816L..19G} 
had an order of magnitude lower mass loading, with  
just a fast, low density component
passing through the $d=1\kpc$ surface where the mass flux is measured.
While this comparison suggests that cosmic rays may be crucial to
accelerating warm outflows, a concern is that the role of SNe may not
have been properly captured in \citet{2016ApJ...816L..19G}.
Due to the high computational cost of including cosmic rays, relatively
low spatial resolution $\Delta x\sim16\pc$ was adopted for this pair of
comparison simulations.
The reported mass outflow rate for SN-only models appears to
differ significantly between the high resolution 
($\Delta x\sim4\pc$) simulation of \citet{2016MNRAS.456.3432G},
with mass flux at $d=1\kpc$ of $4\times10^{-2}\sfrunit$, and the
low-resolution ($\Delta x \sim 16\pc$) simulation of
\citealt{2016ApJ...816L..19G}, with mass flux of $10^{-3}\sfrunit$.
In our own resolution study for a similar parameter regime (see Paper~I),
we found that numerical convergence of SN driven ISM properties 
is not guaranteed at $\Delta x\sim16\pc$, and the low-resolution
outcomes are quite
sensitive to the exact prescription for SN feedback. 
This suggests that higher resolution simulations will be needed
to assess the role of interactions between the cosmic ray fluid and
gas in driving galactic winds.

The \citet{2017MNRAS.466.1903G} models are most similar to our
simulations in that they self-consistently model star formation and SN
feedback, rather than adopting a fixed SN rate.  The main conclusion
they draw is that the outflow rate strongly depends on the volume
filling factor of hot gas.  Above a 50\% volume filling factor, their
measured mass loading factors at $d=1\kpc$ are $\sim 1-100$
(but note that this may be exaggerated since loading is evaluated
instantaneously rather than based on temporal averages; see discussion after
Equation~\ref{eq:energy_loading} in Section~\ref{sec:loading_result}).
While the simulation durations of \citet{2017MNRAS.466.1903G} are
$<100\Myr$, such that star formation and wind mass-loss rates are
likely both subject to ``startup'' transient effects, we agree with
the conclusion that significant driving of fountain flows 
(which dominate at $d=1\kpc$ based on our work) is associated with prominent 
superbubbles near the disk midplane.  In Paper~I, we found the hot
gas fills $\sim 20-60\%$ of the volume at $|z|<H$.  

\citet{2016MNRAS.459.2311M} performed simulations for solar neighborhood
conditions, as well as environments with higher gas surface density and
SN rates.   For high-velocity gas ($\vout>300\kms$) that could potentially
escape, they reported mass loading factors of
0.02-0.005 (lower for higher gas surface density models). 
They also noted the absence of wind acceleration from subsonic
to supersonic velocities as a limitation of local Cartesian box simulations.
We have argued in Section~\ref{sec:wind}
that provided the hot-gas Bernoulli
parameter and mass flux have both approached constant values at large $d$
in a Cartesian simulation, they can be combined to make predictions for
wind properties at large distance.  In this case, the asymptotic
mass flux at high velocity would be larger than the near-disk value
as high-enthalpy gas is further accelerated when streamlines open, but
the total mass flux of hot gas would change little. However, it is not
possible to compare hot-gas mass fluxes
as  \citet{2016MNRAS.459.2311M} did not separate by phase.  

\citet{2017ApJ...841..101L} conducted simulations 
using fixed SN rates, but decomposed by thermal phase in 
reporting mass and energy (as well as metal) loading factors.
Their measurements are at $d=1-2.5\kpc$, with their ``warm'' component
including most of what we consider ``ionized,'' and their ``hot'' extending
down to $T=3\times10^5\Kel$, slightly lower than our ``hot'' definition. 
For their solar neighborhood model, the hot and total 
mass loading factors are about $0.8$ and $2-3$, which are larger than ours by 
a factor of 3 to 8
(see also their discussion in comparison to work of
\citealt{2013MNRAS.429.1922C,2015MNRAS.446.2125C,2016MNRAS.456.3432G}). 
The energy loading factor is also about an order of 
magnitude larger in their model. 

{The reason for this large difference in mass and energy loading
  with respect to our findings is mainly due to the difference in the
  vertical scale height of SNe (relative to the gas scale height), as
  also pointed out in \citet{2017ApJ...841..101L}.  By placing SNe
  randomly with a fixed SN scale height of 250pc, we find that we are
  able to reproduce their results (see
  Appendix~\ref{sec:appendix_sn}). When we adopt a fixed SN
  scale height of 250pc, the majority of SN
  explosions occur outside of the main gas layer. Each SN remnant can then
  expand into the hot, rarefied disk atmosphere with little
  interaction with warm gas.  Most of the injected energy is carried
  outward before cooling.  The
  resulting ISM properties are also quite different: the gas scale height
  is smaller ($H\sim150\pc$), star formation rates are higher, and the
  hot and warm/cold phases are almost completely segregated (single
  phase outflow).  In Appendix~\ref{sec:appendix_sn}, we also test
  models with no runaway O stars, and with 
  random SN placement with a smaller scale height of 50pc.  Both alternatives
  result in loading factors and warm gas velocity distribution very similar to
  the fiducial model.
  Within the standard TIGRESS framework, the spatial distribution of SNe is
subject to the adopted prescription for runaways, but is otherwise 
self-consistently determined with respect to the gas
by the distribution of star formation sites.  
  Better theoretical and observational constraints for runaway OB
  stars would lead to more accurate modeling of the SN distribution,
  which in principle could change the wind mass loading.  However, 
  our tests suggest that a very large proportion of high-velocity
  runaways would be needed to significantly increase the wind mass-loading,
  while the corresponding star formation rate and ISM phase segregation
  in that case might not be consistent with observations.}

Finally, we remark that a fine enough grid to spatially resolve both
low filling-factor warm gas and high filling-factor hot gas in the
wind launching region is crucial for proper physical characterization
of galactic winds.  If warm and hot gas are artificially mixed,
e.g. if the flow in AMR 
and semi-Lagrangian simulations
moves from a higher resolution
region near the midplane to a lower resolution region at high
latitude, the result can be unphysical in ways that would compromise
the implications for real galactic systems.  For example, {consider
mixing of warm and hot flows that have
total horizontally-averaged mass, momentum, and energy vertical fluxes of 
$F_\rho$, $F_{\rho v}$, and $F_E$.  
If we further assume a steady state and neglect
gravity and 
magnetic fields, the
outgoing fluxes of the mixed gas must be the same as the sum of the
horizontally-averaged incoming fluxes of the warm and hot gas. That is,
$F_\rho = \rho_{\rm mix} v_{\rm mix} = \langle \rho_w v_w\rangle + \langle\rho_h v_h\rangle$, and similarly 
$F_{\rho v} = \rho_{\rm mix} v_{\rm mix}^2 + P_{\rm mix}$
and 
$F_{E} = \rho_{\rm mix}v_{\rm mix}
        [v_{\rm mix}^2/2 + \gamma P_{\rm mix}/((\gamma-1)\rho_{\rm mix})]$
are the sum of ``warm'' and ``hot'' terms based on horizontal averages
over multiple zones.}

{
The post-mixing mean velocity will depend on the total fluxes as
\begin{equation}\label{eq:vmix}
v_{\rm mix} = \frac{\gamma}{\gamma+1}\frac{F_{ \rho v}}{F_\rho}
\sbrackets{1-\rbrackets{1-\frac{2(\gamma^2-1)}{\gamma^2}\frac{F_E F_{\rho}}{F_{\rho v}^2}}^{1/2}}.
\end{equation}
Typically, $F_\rho$ (near the disk)
will be dominated by the warm medium contribution (see Fig.
\ref{fig:loading}a), $F_{\rho v}$ will be dominated by the hot medium
contribution 
(see Fig. \ref{fig:zprof}c), and 
$F_E$ will be dominated by the hot medium contribution (see Fig.
\ref{fig:loading}b).
}

\begin{figure}
\plotone{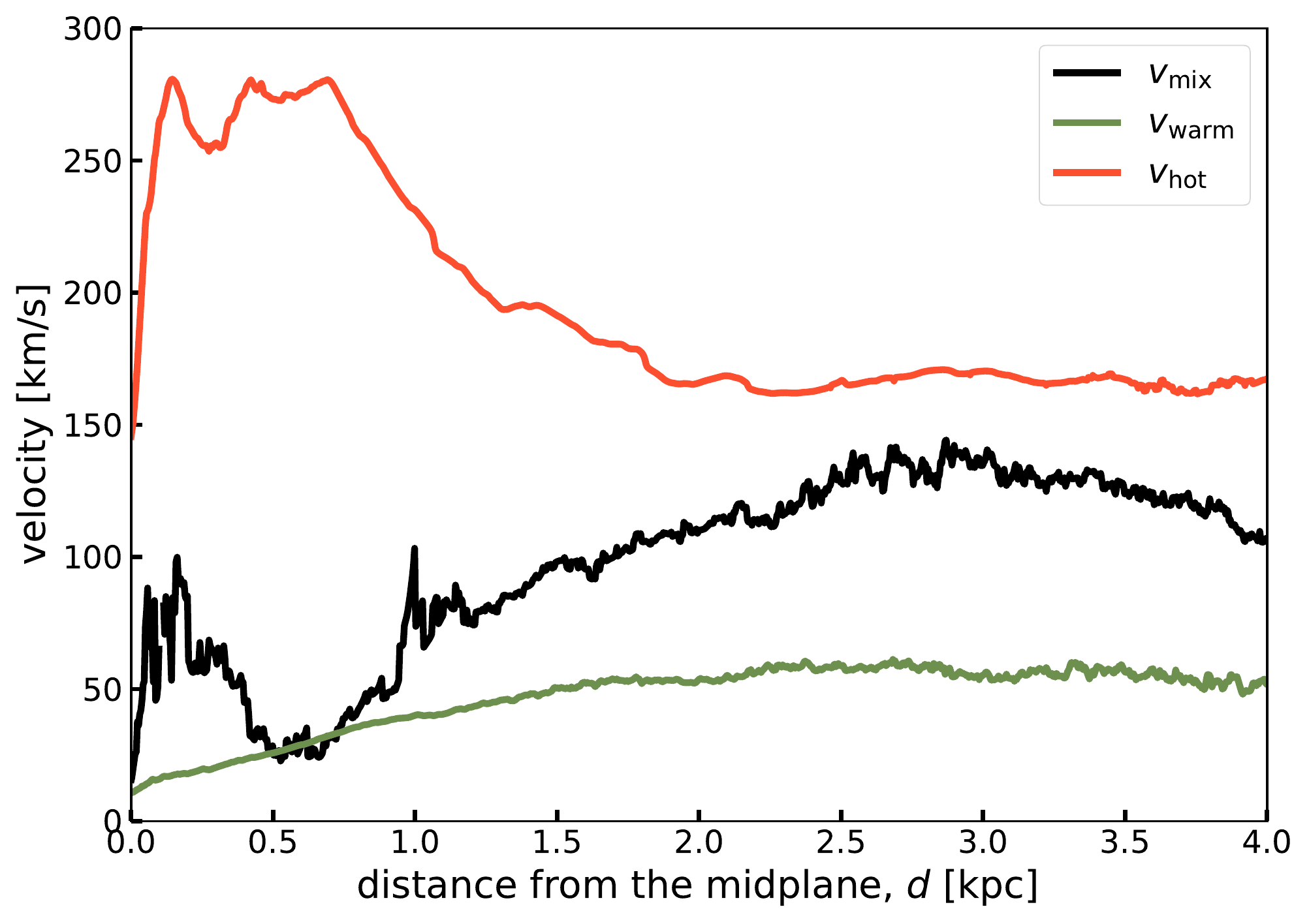}
\caption{The hypothetical post-mixing velocity $v_{\rm mix}$ (black)
  if artificial
  numerical mixing were to happen at a distance $d$ from the midplane.
  We use horizontally-averaged outgoing fluxes to calculate
  $v_{\rm mix}$ using Equation (\ref{eq:vmix}), and take
  a time average over the outflow period.  Also shown are mean values of
  the warm-gas and hot-gas outflowing velocities.  
  The nominal value of $v_{\rm mix}$ is similar to the warm medium velocity
  at small $d$, and gets closer to the hot medium velocity at large $d$.
\label{fig:vmix}}
\end{figure}

{Figure~\ref{fig:vmix} shows this hypothetical
  post-mixing velocity as calculated from 
  our simulation. This is based on horizontal averages of the
  outgoing (i.e. zones with $sign(v_z)=sign(z)$) fluxes, temporally
  averaged over the outflow period and over both sides of the disk.  
Also shown, for comparison, are the mean outflow velocities of the warm and
  the hot gas.
  The implication of the comparison shown in Figure~\ref{fig:vmix}
  is that if numerical mixing were to occur
  at $d>1\kpc$, the result would be
  $v_{\rm mix}\sim100\kms$, which is intermediate between the
  mean hot and mean warm outflow velocities.
}
Depending on the galaxy/halo global
properties, this could have two
(opposite) unphysical consequences. On the one hand, if $v_{\rm mix}$
is greater than the galaxy escape speed but the original $v_w$ is less
than the galaxy escape speed, then the artificially-mixed flow would be
able to escape with a much larger mass-loading factor than would be
realistic (i.e. $\sim \beta_w$ rather than $\beta_h$).  On the other
hand, if $v_{\rm mix}$ is smaller than the galaxy escape speed, then the
artificially-mixed flow would not be able to escape at all, whereas in
reality the hot wind should escape with mass-loading factor $\beta_h$,
and potentially loaded with more than its share of metals.

To avoid these unphysical consequences, it is necessary to separately resolve
the multiphase gas even above the dense midplane region.  

\section{SUMMARY}\label{sec:summary}

Gas flows blown out of galactic midplane regions
by energetic stellar feedback, most
notably type II SNe, will either escape as a wind or turn around
as a fountain, depending on the specific kinetic and thermal energy of the gas 
compared to the gravitational potential.
Due to the multiphase structure of the ISM,
the simplest steady, adiabatic solutions \citep[e.g.,][]{1985Natur.317...44C}
are not directly applicable, although aspects of these solutions are
informative when considering a phase-decomposed analysis.  
More generally, the question of ``how much mass and energy are carried out 
by each phase of outflowing gas?'' depends strongly on when and where
star formation and SNe occur and how feedback transfers energy to the ISM.
Given the complexity involved, the properties of galactic
winds and fountain flows created by the star-forming ISM must
be investigated via fully self-consistent numerical simulations.

The TIGRESS implementation described in Paper~I provides
comprehensive, self-consistent simulations of the multiphase
star-forming ISM that can be used for high-resolution investigations
of outflows in a wide variety of environments.  In this paper, we
analyze the solar neighborhood model of Paper~I to characterize
vertical
gas flows phase-by-phase.
Our principal finding is that the outflowing gas consists of a superposition
of a hot wind and a warm fountain flow.  We summarize key conclusions from
our analysis of each component below.  

\subsection{\it Hot wind}

1. The hot gas component, which is created by SN shocks, behaves very similarly
to expectations for a steady, adiabatic flow as it expands away from
the midplane.  
{For the horizontally-averaged, time-averaged hot component}, the
mass flux and the Bernoulli parameter are close to constant as a 
function of the distance from the midplane $d$ above $d>1\kpc$, indicating
that there is relatively
little mass added or subtracted by heating or cooling from/to
another phase, and little energy added or subtracted either through shocks,
radiation, {pressure work on other phases}, or mixing.

2. The hot gas mass loading factor, defined as the
ratio of outflowing mass flux to the star formation rate per unit area,
is $\beta_h\sim 0.1$, decreasing by
$40\%$ from $d=1$ to $4\kpc$.  The measured value of $\beta_h$ is
consistent with estimates from idealized numerical experiments of
superbubble expansion in a warm-cold ISM.

3. Above $d=1\kpc$, the hot outflow contains a tiny fraction of the
energy originally injected in the ISM by SNe. Another tiny fraction
is converted to kinetic energy of the warm and
cold medium accelerated by superbubble expansion, but most of
the original SN energy is
radiated away within the disk scale height.
The ratio of outflowing hot-gas energy flux to the
SN energy injection rate per unit area is $\alpha_h\sim 0.02$, 
decreasing by 
$65\%$ from $d=1$ to $4\kpc$.
For $d=2-4\kpc$, the mean temperature of the hot gas is $\sim1.5\times10^6\Kel$,
 and its mean velocity is
 $\sim200\kms$. 
Magnetic energy in the outflow is negligible,
with mean Alfv\'en speed of $\sim30\kms$.

4. In the present simulation, the energy
of the hot medium
is carried mainly as heat because the outflow is constrained to remain
vertical in our local Cartesian box.  Allowing for the opening of streamlines
at larger distance (beyond our simulation domain), much of the enthalpy
would be converted to kinetic energy.  The hot medium is barely affected
by gravity within the simulation domain, and with an asymptotic maximum
wind speed of
{$\sim \sqrt{2 ({\cal B} - \Phi)}\sim 350\kms$} 
  would easily escape to very large distances.  
More generally, local simulations with varying galactic conditions
can provide measures of $\beta_h$ and the Bernoulli parameter $\mathcal{B}$
that can be used to provide predictions for hot wind properties
at large distances in a wide range of galaxies.

\subsection{\it Warm Fountain}

1. The warm (including both neutral and ionized) medium, which dominates
the total gas mass in the simulation, acquires energy from SN feedback,
mostly in kinetic form.  However, the typical outgoing velocity of the warm
medium above $d=1\kpc$ is only
$v_{\rm out}\sim 60\kms$, which is not enough to overcome 
the large-scale gravitational potential in the present simulation.
As a result, most of the warm gas that is blown out of the midplane
eventually turns around and falls back, forming a fountain.

2. Star formation bursts lead to a succession of
outflow-dominated and inflow-dominated periods of the fountain flow.
During both outflow-dominated and inflow-dominated periods, gas
with both signs of velocity (i.e. outflows and inflows)
is present on both sides of the disk. Considering just the outflowing gas, 
owing to deceleration and turnaround the 
outgoing warm-medium  mass loading factor $\beta_w$ is a 
\emph{decreasing} function of height $d$.  
The mean warm-gas mass flux in outflowing ``fountain'' gas at $d\sim 1\kpc$ is 
$\beta_w\sim1$, but this drops steeply to $\beta_w\sim0.03$ at $d=4\kpc$.

3. The value of
{the warm-medium energy loading factor} 
$\alpha_w$ drops from $\sim 0.002$ to $10^{-4}$
over $d=1$ to $4\kpc$.  Because $\alpha_w$ drops slightly less than
$\beta_w$, the mean specific energy of the warm medium increases with $d$.
The corresponding
(volume-weighted) 
mean velocities of the warm gas at $d=1$ and $4\kpc$ are
$\sim60$ and $80\kms$.  However, it is important to note that
the increase of mean velocity in the warm medium with $d$ is primarily
due to dropout of low-velocity fluid elements, and \textit{does not} reflect
acceleration with height.
Detailed distributions show a secular decrease in the mass and mass flux 
of high-velocity warm gas with increasing $d$.

4. A promising way to characterize warm outflows is via the velocity
distribution where they are launched at $d\sim H$.  Here, we find that
the high-velocity warm gas has an exponential distribution.  For a given
PDF in velocity, the portion of the warm gas mass flux that is able to
escape as a wind will depend on the halo potential depth.  For the
large scale galactic potential in the present simulation,
very little warm gas escapes, but in 
a dwarf galaxy the same distribution could lead to a wind with 
$\beta_w\sim 1/3$.

\subsection{Caveats and prospects}

Our simulations have two main caveats for direct comparison with observations.
Firstly, we use a local Cartesian box to achieve high resolution.
The uniformly high resolution is crucial for distinguishing different
phases and limiting numerical mixing.  
However, the local Cartesian box prevents us from
following the hot outflow's evolution under global geometry with a  
realistic galactic potential \citep[see][]{2016MNRAS.459.2311M}.
Without the opening of streamlines,
the hot medium cannot accelerate through a sonic point to reach its
asymptotic velocity; we
therefore cannot follow this process directly in our simulations.
However, proper decomposition of the gas phases allows us to provide
well-defined mass and energy loading factors for the hot gas, 
which can be robustly extrapolated to obtain predictions for
asymptotic wind properties at large scales.
The hot-gas loading factors 
{at large $d$ are slightly lower than} the
observational constraints deduced 
from M82 \citep{2009ApJ...697.2030S}.
{This is likely because in M82 the strong starburst has successfully
  cleared much of the cooler gas away from the midplane; we find that when
  SNe explode above the denser phases, the mass and energy loading of hot
  winds increases.}
As discussed above, to the extent that the warm phase is ballistic
above a scale height, its  measured velocity distributions could be used to
extrapolate its properties to large distance.  However, it is clear
that the gravitational potential even within our local box affects the
warm-medium properties at large $d$, so it is important to treat conclusions
regarding warm-medium outflows cautiously.  

Secondly, in the present simulations we do not include photoionization.
This affects the properties of the warm and ionized phases,
for which the photoionization heating
can be important. Inclusion of photoionization is necessary for 
direct comparison with observations, where the line diagnostics are sometimes 
better explained by photoionization rather than a shock model 
\citep[e.g.,][]{2016MNRAS.457.3133C}.  As a first step towards this,
it may be sufficient
to compute ionization in post-processing, as photoionization
is unlikely to be important
to the dynamics of high-velocity warm gas even though it may dominate heating.
Most observations of highly ionized absorption lines are however based on
large apertures, so direct theoretical comparisons would also require
simulations with global geometry.

We are able to run our simulations over an extended period, long after
initial transients that may affect quantitative results reported
by others for
outflows in simulations with similar physics and resolution to
the TIGRESS implementation, but shorter durations.  Other recent
high-resolution simulations do not have self-consistent star formation
and feedback, which may affect outflow loading because this is
sensitive to the spatio-temporal correlation of supernovae with ISM
gas of different phases.  In the future, as more groups run
high-resolution simulations with self-consistent star formation and SN
feedback for an extended duration, it will be informative to
compare phase-separated results for mass and energy loading, Bernoulli
parameters, and velocity distributions, for both fountain flows and
winds.

Application of the TIGRESS implementation to other galactic
environments is currently underway, and promises to be quite interesting.
Varying the basic model
input parameters (especially gas and stellar surface density and
metallicity) will enable predictions for multiphase outflow properties
in a wide range of galaxies, and will provide detailed information needed
to build subgrid models for winds in cosmological simulations of galaxy
formation.  By extending TIGRESS and other self-consistent multiphase
ISM/star formation numerical implementations to include additional feedback
(especially radiation and cosmic rays), understanding the complex physics
behind galactic winds and fountains is within reach.

\acknowledgements

{We are grateful to the referee for helpful report, and to
Miao Li and Drummond Fielding for fruitful discussions.}  
This work was supported by grants AST-1312006 from the National Science
Foundation, and NNX14AB49G and NNX17AG26G from NASA.
Resources supporting this work were provided in part by the NASA
High-End Computing (HEC) Program through the NASA Advanced
Supercomputing (NAS) Division at Ames Research Center and in part by
the Princeton Institute for Computational Science and Engineering
(PICSciE) and the Office of Information Technology's High Performance
Computing Center.

\software{This work made use of the {\tt Athena} MHD code \citep{2008ApJS..178..137S,2009NewA...14..139S}. 
This work also made use of analysis and visualization softwares including {\tt yt \citep{2011ApJS..192....9T}, 
astropy \citep{2013A&A...558A..33A}, matplotlib \citep{Hunter:2007}, numpy \citep{vanderWalt2011}, 
IPython \citep{Perez2007}, pandas \citep{mckinney-proc-scipy-2010}}, and {\tt VisIt} \citep{HPV:VisIt}.}

\bibliographystyle{aasjournal} 
\bibliography{ms_cleaned.bbl}{}

\appendix
\section{MHD equations in a Shearing-Box}\label{sec:appendix}

{The ideal MHD equations of mass, momentum, and total energy conservation}
in frame rotating at $\mathbf{\Omega}=\Omega\zhat$
are
\begin{equation}\label{eq:a_mass}
{\pderiv{\rho}{t}+}
\divergence{\rho\vel}=0,
\end{equation}
\begin{equation}\label{eq:a_mom}
{\pderiv{(\rho\vel)}{t}+}
\divergence[\sbrackets]{\rho\vel\vel+P+\frac{B^2}{8\pi}-\frac{\Bvec\Bvec}{4\pi}}=
-2\mathbf{\Omega}\times(\rho\vel)-\rho\nabla\Phi_{\rm tot},
\end{equation}
and
\begin{equation}\label{eq:a_energy}
{\pderiv{\mathcal{E}}{t}+}
\divergence[\sbrackets]{\rbrackets{\frac{1}{2}v^2 +
\frac{\gamma}{\gamma-1}\frac{P}{\rho}+\Phi_{\rm tot}}\rho\vel
+\frac{(\Bvec\times\vel)\times\Bvec}{4\pi}}=
-\rho\mathcal{L},
\end{equation}
where the total energy density is
\begin{equation}
\mathcal{E}\equiv\frac{1}{2}\rho v^2 + \frac{P}{\gamma-1}+\frac{B^2}{8\pi}.
\end{equation}
The total gravitational potential 
$\Phi_{\rm tot}$ may include self-gravity and external-gravity, as well as 
the tidal potential $\Phi_{\rm tidal}=-q\Omega^2x^2$ (see Paper~I).
Note that the zero point of $\Phi_{\rm tot}$ is at the center of the simulation
domain.

With the help of shearing-periodic boundary conditions
in the horizontal directions,
the horizontally-averaged equations are then given by
\begin{equation}\label{eq:a_massz}
{\pderiv{\abrackets{\rho}}{t}+}
\pderiv{}{z}\abrackets{\rho v_z}=0,
\end{equation}
\begin{equation}\label{eq:a_momz}
{\pderiv{\abrackets{\rho v_z}}{t}+}
\pderiv{}{z}\abrackets{\rho v_z^2+P+\frac{B^2}{8\pi}-\frac{B_z^2}{4\pi}}=
-\abrackets{\rho\pderiv{\Phi}{z}},
\end{equation}
and
\begin{equation}\label{eq:a_energyz}
{\pderiv{\abrackets{\mathcal{E}}}{t}+}
\pderiv{}{z}\abrackets{\rho v_z\mathcal{B}+\mathcal{S}_z}=\frac{q\Omega}{L_y}\int\rbrackets{\rho v_x \delta v_y - \frac{B_xB_y}{4\pi}}dy -\abrackets{\rho\mathcal{L}},
\end{equation}
where the angle brackets denote a horizontal average,
$\abrackets{q}\equiv \sum qdxdy/L_xL_y$.
Here, specific energy of the gas,
or the Bernoulli parameter, is defined by
\begin{equation}\label{eq:a_Bernoulli}
\mathcal{B}\equiv \frac{v^2}{2}+\frac{\gamma}{\gamma-1}\frac{P}{\rho}+\Phi,
\end{equation}
where $\Phi=\Phi_{\rm tot}-\Phi_{\rm tidal}$,
and the Poynting vector is defined by
\begin{equation}\label{eq:a_Poynting}
\mathbf{S}\equiv \frac{(\Bvec\times\vel)\times\Bvec}{4\pi},
\end{equation}
with its vertical component of
\begin{equation}\label{eq:a_Poynting_z}
S_z= \frac{v_z B^2 - B_z \vel\cdot\Bvec}{4\pi}.
\end{equation}
The right hand side of Equation~\ref{eq:a_energyz} is the sum of 
energy source terms due to
the Reynolds and Maxwell stresses (on the $x$-surfaces of the box,
from shearing boundary conditions --
e.g., \citealt{1996ApJ...464..690H,2010ApJS..189..142S})
and the net cooling.

{Although the evolution of our simulation is highly dynamic and
  time-dependent, the fluctuations in the 
  horizontally-averaged gas properties are
  with respect to a well-defined equilibrium state
(see Figure~\ref{fig:zprof}), after an 
  early transient period.
We thus take time averages to consider a quasi-steady equilibrium state, 
dropping the time derivative terms in Equations (\ref{eq:a_massz})-(\ref{eq:a_energyz})
to analyze characteristics of mean gas flows.}
We explicitly measured each energy source term
{for separate thermal phases} and 
confirmed that for the hot gas the energy source terms are negligible,
and for the warm gas the cooling dominates the energy source terms. The hot gas
can thus be treated nearly adiabatically. 
{Above the region where SN energy is injected,  
the hot gas mass and energy fluxes remain (roughly) constant,
and hence the ratio of the two}, 
$\mathcal{B} + \mathcal{M}_z$, is also (roughly) constant.  Here
\begin{equation}\label{eq:spec_mag}
\mathcal{M}_z\equiv \frac{S_z}{\rho v_z}
\end{equation}
is the specific magnetic energy carried by the magnetic field,
{ which is shown to be small compared to $\mathcal{B}$
  (Figure~\ref{fig:hot}).  Thus, outside of
  the energy injection region where the hot component is created, its
  mean vertical mass flux $\rho v_z$ and its Bernoulli parameter $\cal B$ are
  expected to be roughly constant,
  independent of distance relative to the midplane.  }

{We note that more generally, if we consider the time-averaged
  state of a flow (i.e. $\partial(...)/\partial t=0$) that is weakly magnetized
  and has negligible cooling, Equations (\ref{eq:a_mass}) and
  (\ref{eq:a_energy}) reduce to $\nabla \cdot (\rho \mathbf{v})=0$ and
  $\mathbf{v} \cdot \nabla {\cal B}=0$.  That is, the Bernoulli parameter
  is conserved along streamlines, regardless of the geometry of the flow.}

\section{Numerical Convergence Tests}\label{sec:appendix_conv}
{ Due to complex and nonlinear interactions between SN feedback
  and the surrounding medium, satisfaction of simple physical
  conditions (e.g., resolving the Sedov stage of a single SN blast wave;
  \citealt{2015ApJ...802...99K}) will not guarantee numerical
  convergence of the simulated system as a whole.  It is therefore
  important to check robustness of our main results for varying
  numerical resolution.}

{In addition, it has been suggested
  that the SN placement can significantly impact the resulting ISM
  structure and feedback efficiency
  \citep[e.g.,][]{2015MNRAS.449.1057G,2015ApJ...814....4L}.
In TIGRESS, unlike in most other simulations, the 
  SN locations are not set based on a pre-defined vertical
  distribution, but are self-determined based on the locations where
  star clusters form and migrate over time.  A fraction of the SNe in our
  simulation are associated with runaways, and we adopt a distribution
  for the ejection velocity from clusters based on a binary population
  synthesis model \citep{2011MNRAS.414.3501E}.  This distribution,
  together with the fraction of
  binary runaways, is in fact not very certain.  It is therefore valuable
  to test how our results might depend on the placement of SNe 
by comparing different
  test runs.  }

\subsection{Numerical Resolution}
\begin{figure}
\plotone{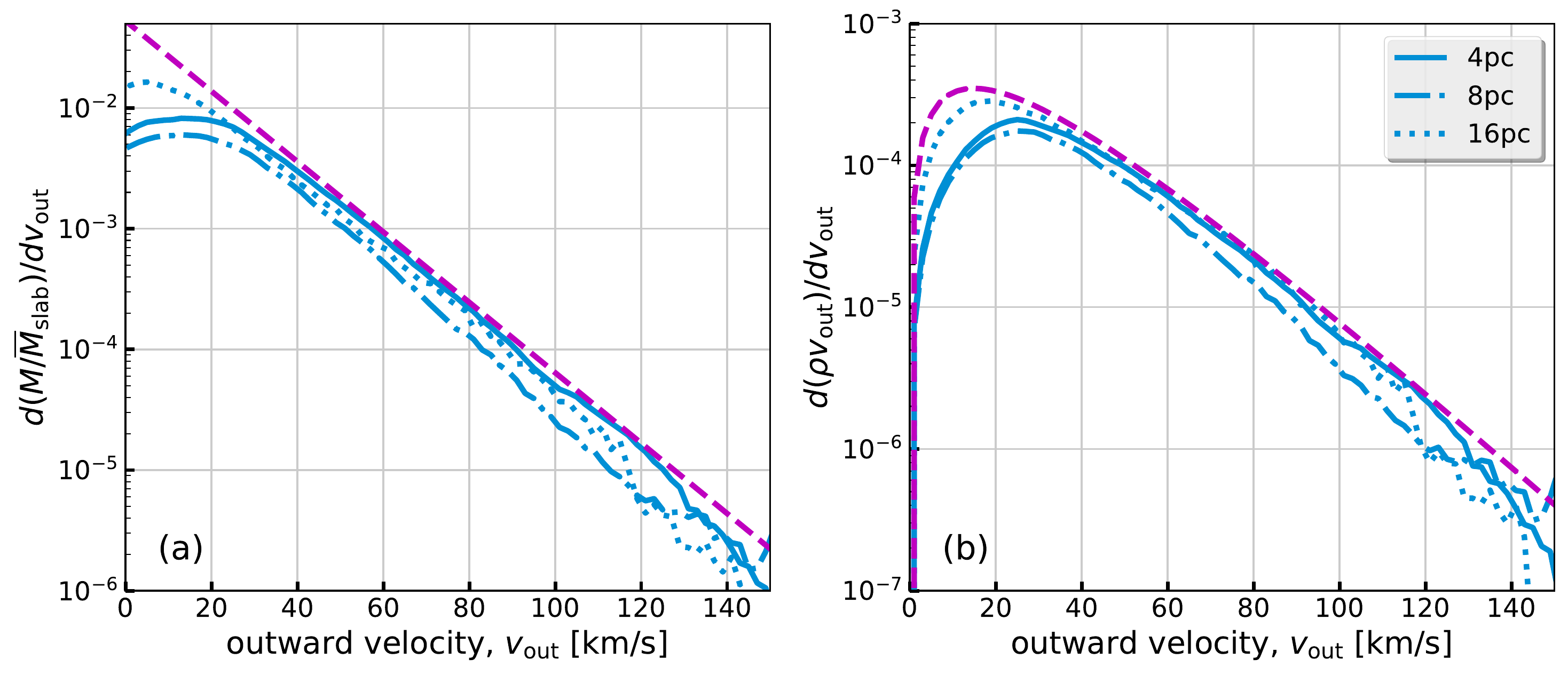}
\caption{Mass- and mass flux-weighted velocity distributions
of the warm medium measured at $d=1\kpc$ for different numerical resolutions
during the outflow period (same as Figure~\ref{fig:warm_vpdf}(c) and (e)).
The magenta dashed lines in each panel are fits given by Equations (\ref{eq:fit}) and (\ref{eq:fit_mf}).
\label{fig:vpdf_conv_res}}
\end{figure}

\begin{figure}
\plotone{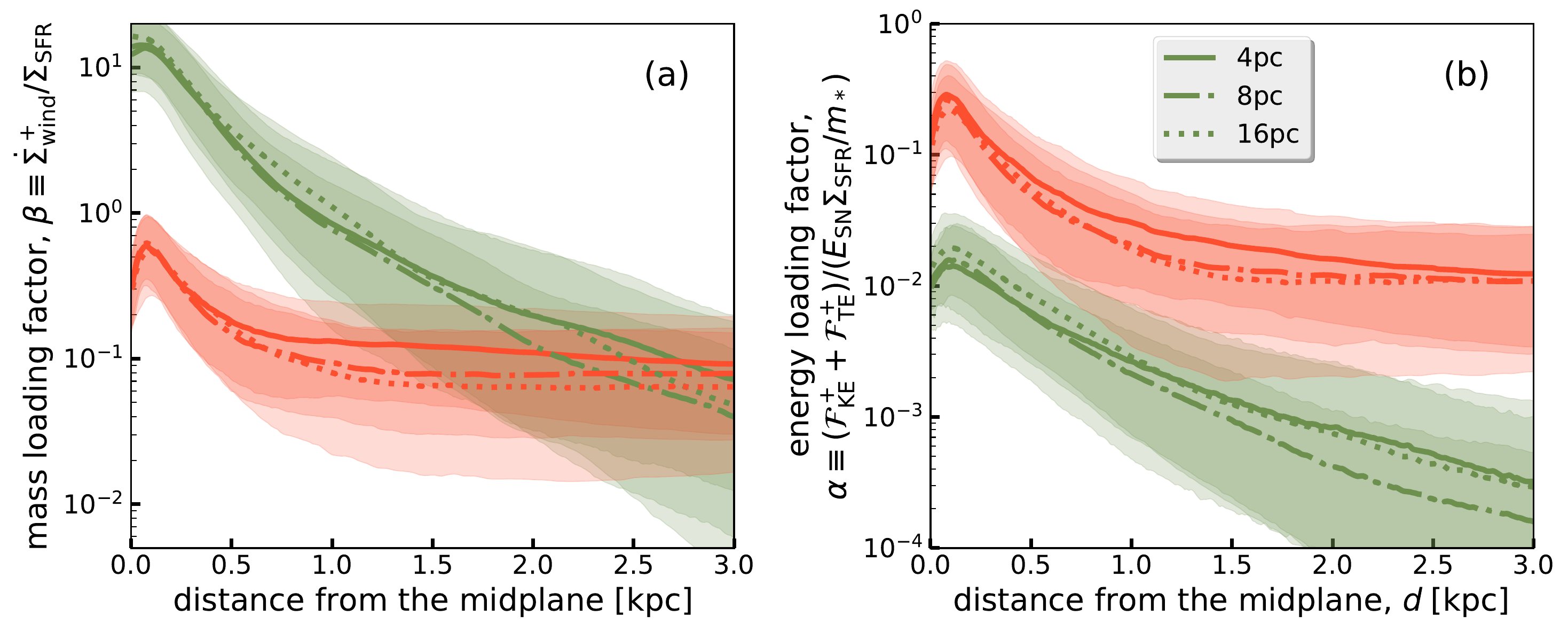}
\caption{Mass and energy loading factors of the warm (green) and hot (red) phases (same as Figure~\ref{fig:loading}),
for different numerical resolutions.
\label{fig:loading_conv_res}}
\end{figure}

{
We analyze the same set of simulations with different numerical resolutions of $\Delta x=$4, 8, and 16 pc. Note 
that we showed in Paper I that
$\Delta x=16\pc$ is a marginal condition for convergence in star formation rates
 and the ISM properties. Here, we further
show detailed velocity distributions of the warm medium (Figure~\ref{fig:vpdf_conv_res}) and mass and energy
loading factors of the warm and hot medium (Figure~\ref{fig:loading_conv_res})
at varying resolution.
These properties, which are key characteristics of 
hot winds and warm fountains, are evidently converged.
}

\subsection{SN Placement}\label{sec:appendix_sn}
{
  In order to compare the effects of SN placement,
  we consider, in addition to the fiducial model,
three models with different prescriptions: (1) {\tt no-runaway} --  all SNe are in star cluster particles without runaways;
(2) {\tt random-250} -- all SNe are randomly located horizontally and 
consistent with an exponential vertical profile 
with scale height of $z_{\rm SN}=250\pc$ vertically,
but with the rate determined by star cluster particles;
(3) {\tt random-50} -- the same as {\tt random-250}, but $z_{\rm SN}=50\pc$. 
}

\begin{figure}
\plotone{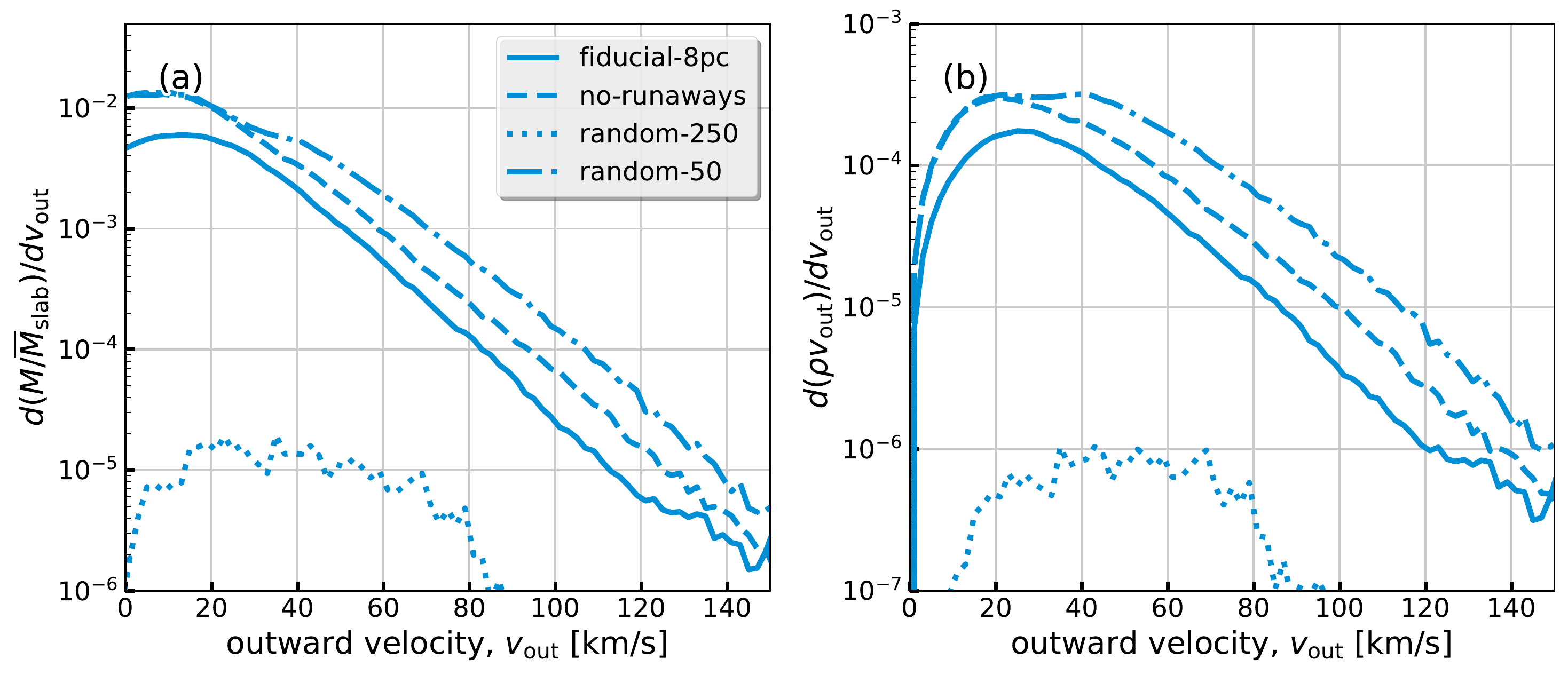}
\caption{Mass- and mass flux-weighted velocity distributions
of the warm medium measured at $d=1\kpc$ for different SN prescriptions
during the outflow period (same as Figure~\ref{fig:warm_vpdf}(c) and (e)).
\label{fig:vpdf_conv_sn}}
\end{figure}

\begin{figure}
\plotone{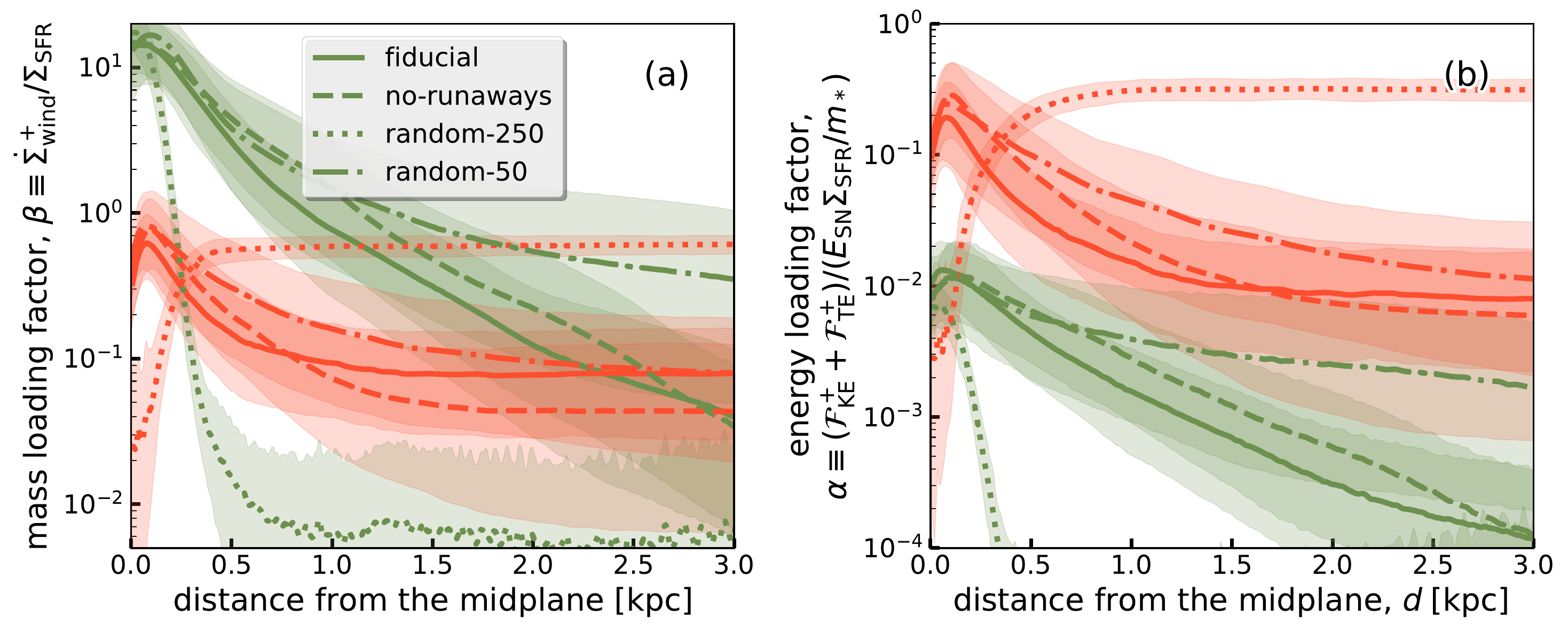}
\caption{Mass and energy loading factors of the warm (green) and hot (red) phases
for different SN prescriptions (same as Figure~\ref{fig:loading}).
\label{fig:loading_conv_sn}}
\end{figure}

{
  By comparing the fiducial model with the {\tt no-runaway} model for
  velocity distributions of the warm medium
  (Figure~\ref{fig:vpdf_conv_sn}) and
mass and energy loading factors of the warm and hot medium (Figure~\ref{fig:loading_conv_sn}),
we conclude that presence or absence of runaways with our adopted prescription
does not significantly alter the results. 
}

{
  With randomly located SNe,
  the results can change
  substantially depending on the adopted scale height. For the {\tt
    random-250} model, the hot gas mass and energy loading factors are
  about $0.8$ and $0.2$, respectively. In this case, most SNe 
  explode above the gaseous scale height without interacting with the
  warm and cold medium. The hot SN remnants simply expand outward in the
  low-density, hot atmosphere.  A negligible warm fountain is created.
  In contrast, for the {\tt
    random-50} model, the majority of SN events happen within the warm/cold gas
  layer.  The hot gas created in SN remnants strongly interacts with the
  surrounding warm and cold medium, and hot gas is lost in the process.
  The resulting hot gas mass and energy loading factors in the wind
  for the {\tt  random-50} model are
  similar to the fiducial model. This trend with varying SN scale height
  was also shown in
  \citet{2017ApJ...841..101L}.  }

{ In short, the mass and energy loading factors of the hot wind
  are sensitive to the vertical distribution of SNe relative to the
  gas (mainly the warm medium). In our fiducial model, we assume
  $2/3$ of SNe explode in binaries and eject runaway companions.
  This gives a runaway SNe fraction of 1/3. Since the ejection
  velocity distribution is an exponential function with
  characteristic velocity $50\kms$, the fraction of SNe that explode
  above $200\pc$ is 15\%. This fraction increases to 44\% in the {\tt
    random-250} model, while the {\tt no-runaway} and {\tt random-50}
  models give fractions of 0.4\% and 7\%, respectively.  }
 
{ It is noteworthy that the {\tt random-250} model fails to
  regulate star formation rates at the same level as the fiducial
  model (the SFR is about 1.5 times higher for the {\tt random-250}
  model).  In the {\tt random-50} model, an asymmetric vertical
  distribution of gas develops at later times (more gas is in the
  upper half of the simulation domain). Since the SNe distribution is set
  relative to the initial gas distribution rather than the instantaneous
  gas distribution, the asymmetric gas structure persists and allows
  more efficient hot and warm outflows in
  the lower half of the domain. This results in slightly higher mass and energy
  loading factors of the warm fountain for the {\tt random-50} model relative
  to the fiducial model.
}

\end{document}